\documentclass[prd,12pt]{article}

\usepackage{amsmath,amssymb,graphicx,multirow,xspace,slashed,array,booktabs}
\usepackage[colorlinks=true,urlcolor=blue,anchorcolor=blue,citecolor=blue,filecolor=blue,linkcolor=blue,menucolor=blue]{hyperref}
\usepackage[compress,numbers]{natbib}
\usepackage[labelformat=simple]{subcaption}

\usepackage{booktabs}
\usepackage{blindtext, rotating}
\usepackage{afterpage}
\usepackage{enumitem}
\usepackage{marvosym}
\usepackage{authblk}
\usepackage{verbatim}
\usepackage{multirow}
\usepackage{soul} 
\usepackage[normalem]{ulem}
\usepackage{pifont}
\usepackage{booktabs}
\usepackage{bbm}

\usepackage{floatrow}
\newfloatcommand{capbtabbox}{table}[][\FBwidth]

\usepackage[font=footnotesize,labelfont=bf]{caption}

\usepackage{lineno}

\allowdisplaybreaks

\long\def\symbolfootnote[#1]#2{\begingroup%
\def\thefootnote{\fnsymbol{footnote}}\footnote[#1]{#2}\endgroup}

\allowdisplaybreaks

\makeatletter
	
	\@addtoreset{equation}{section}
\makeatother

\newcommand{\newc}{\newcommand}
\newc{\gsim}{\lower.7ex\hbox{$\;\stackrel{\textstyle>}{\sim}\;$}}
\newc{\lsim}{\lower.7ex\hbox{$\;\stackrel{\textstyle<}{\sim}\;$}}
\newc{\gev}{\,{\rm GeV}}
\newc{\mev}{\,{\rm MeV}}
\newc{\ev}{\,{\rm eV}}
\newc{\kev}{\,{\rm keV}}
\newc{\tev}{\,{\rm TeV}}
\newc{\MHT}{$H_T^{\text{miss}}$}
\newc{\MET}{$\slashed{E}_T$}
\newc{\MTT}{$M_{T2}$}

\newc{\mz}{M_Z}
\newc{\mpl}{M_*}
\newc{\mw}{m_{\rm weak}}
\newc{\nr}[1]{N^c_R{}_{#1}}

\def\beq{\begin{equation}}
\def\eeq{\end{equation}}
\newcommand{\bea}{\begin{eqnarray}\begin{aligned}}
\newcommand{\eea}{\end{aligned}\end{eqnarray}}
\def\bitem{\begin{itemize}}
\def\eitem{\end{itemize}}

\begin{document}
\baselineskip 0.6cm

\begin{titlepage}

\vspace*{-1.5cm}

\thispagestyle{empty}

\begin{flushright}
KYUSHU-HET-347
\end{flushright}

\begin{center}
\vskip 1cm

{\Large \bf The Minimal Supersymmetric Standard Model \\[0.5ex] with Non-Invertible Selection Rules}

\vskip 1cm
{\large Yuichiro Nakai$^{1,2}$, Hajime Otsuka$^{3,4}$, \\[1ex]
Yoshihiro Shigekami$^{5}$ and Zhihao Zhang$^{1,2}$}
\vskip 0.5cm
{\it
$^1$Tsung-Dao Lee Institute, Shanghai Jiao Tong University, \\ No.~1 Lisuo Road, Pudong New Area, Shanghai 201210, China \\
$^2$School of Physics and Astronomy, Shanghai Jiao Tong University, \\ 800 Dongchuan Road, Shanghai 200240, China \\
$^3$Department of Physics, Kyushu University, \\ 744 Motooka, Nishi-ku,  Fukuoka 819-0395, Japan \\
$^4$Quantum and Spacetime Research Institute (QuaSR), Kyushu University, \\ 744 Motooka, Nishi-ku, Fukuoka 819-0395, Japan\\
$^5$School of Physics, Henan Normal University, Xinxiang, Henan 453007, China}
\vskip 1.0cm

\end{center}

\begin{abstract}

\vskip 0.3cm
We investigate a framework of the Minimal Supersymmetric Standard Model (MSSM) in which the quark and lepton flavor structure and suppression of flavor-changing neutral currents (FCNCs) are governed by non-invertible selection rules. 
By implementing such non-group-like fusion rules for matter fields, arising from gauging the outer automorphism $\mathbb{Z}_2$ of a discrete $\mathbb{Z}_N$ symmetry, we obtain realistic Yukawa textures that reproduce the observed quark and lepton masses and mixings while ensuring diagonal soft supersymmetry (SUSY) breaking masses and hence suppressing dangerous FCNC processes. 
We analyze mass insertion parameters under random $\mathcal{O}(1)$ coefficients and find that all flavor-violating effects are consistent with experimental limits on processes such as $\mu \to e \gamma$ and meson mixings. 
We show that the Yukawa textures and soft terms remain stable under renormalization group evolution. 
Our results demonstrate that non-invertible selection rules provide a compelling new mechanism to address both the flavor structure and FCNC problems in supersymmetric models. 

\end{abstract}

\flushbottom

\end{titlepage}


\section{Introduction}
\label{intro}

One of the central questions in modern particle physics is the origin of masses of fundamental particles. 
Within the Standard Model (SM), the Higgs mechanism provides a successful description of mass generation, yet it leaves two fundamental issues unresolved. 
First, the pattern of masses and mixings among the three generations of quarks and leptons is highly nontrivial and remains unexplained. 
The SM offers no underlying principle accounting for why this specific structure is realized in nature. 
Second, the Higgs field is the only elementary scalar in the SM, and its mass is extremely sensitive to ultraviolet physics, requiring severe fine-tuning to remain at the electroweak scale. 
These open questions have motivated extensive theoretical and experimental efforts to explore physics beyond the SM, yet a definitive resolution has remained elusive. 

In the program to uncover the underlying principle for the pattern of SM fermion masses and mixings, recently, \emph{non-invertible selection rules} have emerged as a powerful organizing principle capable of overcoming limitations of the ordinary group-based symmetries. 
The non-invertible selection rules arising from non-invertible fusion algebras impose constraints on Yukawa couplings that cannot be captured by group-based symmetries~\cite{Kobayashi:2024yqq,Kobayashi:2024cvp}. 
The obtained Yukawa textures successfully reproduce the masses and mixings of quarks~\cite{Kobayashi:2025znw} and leptons~\cite{Kobayashi:2025ldi}.\footnote{Non-invertible selection rules may also address the strong CP problem~\cite{Cordova:2024ypu,Liang:2025dkm,Kobayashi:2025thd,Kobayashi:2025rpx}. 
For other phenomenological applications, see, e.g., Refs.~\cite{Suzuki:2025oov,Kobayashi:2025cwx,Kobayashi:2025lar,Nomura:2025yoa,Chen:2025awz,Okada:2025kfm,Nomura:2025tvz}.}
Such selection rules appear in theories with non-invertible symmetries in various dimensions (see, e.g., Refs.~\cite{Schafer-Nameki:2023jdn,Brennan:2023mmt,Bhardwaj:2023kri,Luo:2023ive,Shao:2023gho} for reviews).\footnote{There have been remarkable developments in applying non-invertible symmetries to particle phenomenology, e.g., the emergence of non-invertible symmetries from Abelian chiral symmetries~\cite{Choi:2022jqy,Cordova:2022ieu} and a realization of small Dirac neutrino Yukawa couplings protected by a non-invertible symmetry~\cite{Cordova:2022fhg}.}
Moreover, it has turned out that the non-invertible selection rules are motivated by string compactifications, as discussed in type IIB magnetized D-brane models~\cite{Kobayashi:2024yqq}, heterotic string theory on Calabi-Yau threefolds~\cite{Dong:2025pah}, and Abelian and non-Abelian orbifolds~\cite{Dong:2025jra,Kobayashi:2025ocp}. 
Therefore, non-invertible selection rules provide a promising and theoretically grounded path toward understanding the origin of Yukawa structures. 

Supersymmetry (SUSY) has long been regarded as one of the most compelling solutions to the electroweak naturalness problem of the SM (see e.g. Ref.~\cite{Martin:1997ns} for a review). 
By pairing bosons and fermions, SUSY cancels quadratic divergences in scalar masses and thereby stabilizes the electroweak scale against radiative corrections. 
The Minimal Supersymmetric Standard Model (MSSM), as the simplest supersymmetric extension of the SM, provides a concrete framework in which this cancellation mechanism is realized, offering a well-motivated setting for physics at and beyond the TeV scale. 
However, if SUSY is realized in nature, it must be broken, and the associated soft SUSY breaking terms generically introduce new sources of flavor violation. 
For instance, the squark and slepton mass matrices in the MSSM are not necessarily aligned with the SM Yukawa matrices. 
After rotating to the fermion mass basis, this misalignment leads to sizable flavor-changing neutral currents (FCNCs), potentially exceeding experimental bounds from $K^0$--$\overline{K^0}$ mixing, $B$-meson oscillations, and charged-lepton flavor-violating processes. 
This tension constitutes the SUSY flavor problem. 
Among various solutions, the \emph{alignment mechanism} arranges the flavor structure of the soft SUSY breaking terms to be approximately diagonal in the same basis that diagonalizes the Yukawa matrices~\cite{Leurer:1992wg,Nir:1993mx,Leurer:1993gy,Ibanez:1994ig,Nir:1996am,Grossman:1995hk,Aloni:2021wzk,Nakai:2021mha}. 
When realized, this alignment suppresses sfermion mixing angles and ensures that SUSY contributions to FCNCs remain within observational limits. 
Constructing models that naturally enforce such alignment is therefore essential for achieving a phenomenologically viable supersymmetric particle spectrum. 

In the present paper, we utilize a non-invertible selection rule realized by $\mathbb{Z}_2$ gauging of a $\mathbb{Z}_N$ symmetry~\cite{Kobayashi:2024yqq,Kobayashi:2024cvp} to find viable Yukawa textures in the MSSM framework. 
Assuming that SUSY breaking also respects the non-invertible selection rule, the alignment mechanism works. 
In particular, the soft scalar masses are diagonalized in a flavor basis, which leads to a significant suppression of SUSY contributions to FCNCs. 
We explicitly demonstrate that the Yukawa textures reproducing realistic quark and lepton mass matrices, realized by the non-invertible selection rule, lead to a solution to the SUSY flavor problem in the MSSM. 
It has been pointed out in Refs.~\cite{Heckman:2024obe,Kaidi:2024wio,Funakoshi:2024uvy} that non-invertible selection rules are in general broken by quantum corrections.\footnote{See Refs.~\cite{Suzuki:2025bxg,Suzuki:2025kxz} for spurion analysis of non-invertible selection rules.}
We then discuss quantum corrections to the non-invertible selection rule by taking account of renormalization group effects on the MSSM parameters. 
It is shown that the Yukawa textures and soft SUSY breaking terms are sufficiently stable under renormalization group evolution from a high energy scale down to the soft mass scale. 
Our results illustrate that non-invertible selection rules offer a robust and attractive framework for simultaneously addressing the origin of flavor structures and ensuring flavor safety in the MSSM. 

The rest of the paper is organized as follows. 
In section~\ref{model}, we introduce our non-invertible selection rule into the MSSM and present the Yukawa textures and soft SUSY breaking terms at a UV scale, which give initial conditions for renormalization group flows to the soft mass scale. 
Then, in section~\ref{FCNC}, we analyze SUSY contributions to FCNC processes in our model and show that FCNC constraints are satisfied with TeV-scale superpartners. 
Section~\ref{conclusion} is devoted to conclusions and discussions. 
Appendix~\ref{app} summarizes Yukawa textures for quarks other than those presented in the main text.

\section{Setup}
\label{model}

We begin to introduce our non-invertible selection rule and apply it to the MSSM. 
The selection rule leads to specific Yukawa textures and constrains structures for soft SUSY breaking terms at a UV scale where the rule is applied. 
Then, we discuss the stability of Yukawa textures under renormalization group evolution and show soft SUSY breaking parameters at the TeV scale.

\subsection{The selection rule}
\label{sec:setup}

We assume that fields in our four-dimensional theory obey a non-invertible fusion rule obtained by gauging the outer automorphism of a group (see, e.g., Refs.~\cite{Bartsch:2022mpm,Bhardwaj:2022yxj}). 
Let us focus on the $\mathbb{Z}_2$ gauging of a $\mathbb{Z}_N$ symmetry~\cite{Kobayashi:2024yqq,Kobayashi:2024cvp}. 
For a generator of $\mathbb{Z}_N$, $g$, we consider the following $\mathbb{Z}_2$ outer automorphism:
\begin{align}
r^2=e\,,\qquad r g^k r^{-1} = g^{-k} \, ,
\end{align}
where the outer automorphism of $\mathbb{Z}_N$ is represented by $r$, mapping an element of $\mathbb{Z}_N$ $g^k$ to its inverse $g^{-k}$. Here, $e$ denotes the identity. Then, one can define classes:
\begin{align}
[g^k] = \{ h g^k h^{-1} | h = e, r \} = \{ g^k, g^{-k} \} \, ,
\label{eq:class}
\end{align}
with $k = 0, 1, ..., \lfloor \frac{N}{2} \rfloor$, where $\lfloor \frac{N}{2} \rfloor$ denotes the floor function. 
The above equivalence classes correspond to $\mathbb{Z}_2$ invariant $D_N \cong \mathbb{Z}_N \rtimes \mathbb{Z}_2$ conjugacy classes, motivated by string compactifications such as type IIB string theory on toroidal orbifolds~\cite{Kobayashi:2024yqq}.\footnote{Other classes of non-invertible selection rules are also reported in Refs.~\cite{Dong:2025pah,Dong:2025jra}.}
In magnetized compactifications, there exists a $\mathbb{Z}_N$ flavor symmetry for four-dimensional massless modes obtained through the Kaluza-Klein reduction of a higher-dimensional Yang-Mills theory on a torus~\cite{Abe:2009vi,Berasaluce-Gonzalez:2012abm,Marchesano:2013ega}. 
However, it is broken by the $\mathbb{Z}_2$ orbifold twist, and then $\mathbb{Z}_2$-even modes $\phi^k$ are labeled by a class $[g^{k}]$ in Eq.~\eqref{eq:class} obeying~\cite{Kobayashi:2024yqq}
\begin{align}
[g^{k_1}] \otimes [g^{k_2}] = [g^{k_1+k_2}] \oplus [g^{k_1-k_2}] \, .
\end{align}
Since each field is labeled by a class, selection rules are different from ordinary group-like ones. 
Indeed, when we denote the element of a class $[g^k]$ as $\tilde{g}^k$, a bare coupling of four-dimensional fields $\phi_{k_1} \cdots \phi_{k_n}$, each of which is labeled by a class $[g^{k_i}]$, is allowed when there exist elements $\tilde{g}^{k_i} \in [g^{k_i}]$ satisfying
\begin{align}
\label{eq:cond}
\tilde{g}^{k_1} \cdots \tilde{g}^{k_n} = e \, .
\end{align}
In a model with group-theoretical symmetries, each field carries a unique, fixed charge, and a coupling is allowed if and only if the sum of these unique charges is zero. In contrast, under the non-invertible selection rule in Eq.~\eqref{eq:cond}, a field corresponds to a set of elements. A coupling is allowed if at least one combination of elements chosen from the respective classes satisfies the conservation law. 
Note that the conjugate of $\phi_{k_i}$ is also labeled by the same class $[g^{k_i}]$. 
These non-invertible selection rules can be formulated in terms of hypergroups, in particular fusion algebras, and these mathematically well established concepts are known to be consistent with quantum field theory and string theory (see, e.g., Ref. \cite{Kaidi:2024wio}). 

Let us now consider the MSSM with three right-handed neutrinos. 
The matter fields obey a non-invertible selection rule realized by a $\tilde{\mathbb{Z}}_5^{(1)} \times \tilde{\mathbb{Z}}_5^{(2)}$ symmetry, where we denote $\tilde{\mathbb{Z}}_5$ for the $\mathbb{Z}_2$ gauging of a $\mathbb{Z}_5$ in what follows. 
For each $\tilde{\mathbb{Z}}_5$ symmetry, there exist three independent classes $\{[g^0], [g^1], [g^2]\}$ obeying
\begin{align}
[g^0] \otimes [g^0] &= [g^0] \, , \nonumber \\
[g^0] \otimes [g^j] &= [g^j] \otimes [g^0] = [g^j] \, , \nonumber \\
[g^1] \otimes [g^1] &= [g^0] \oplus [g^2] \, , \nonumber \\
[g^1] \otimes [g^2] &= [g^2] \otimes [g^1] = [g^1] \oplus [g^2] \, , \nonumber \\
[g^2] \otimes [g^2] &= [g^0] \oplus [g^1] \, ,
\end{align}
with $j = 1, 2$. 
We choose the assignment of the matter superfields as
\begin{align}
Q_{i(=1,2,3)} &: ([g^2], [g^1], [g^2]) \, , \qquad U_i: ([g^1], [g^0], [g^1]) \, , \qquad D_i: ([g^0], [g^2], [g^2]) \, , \nonumber \\
L_i &: ([g^0], [g^1], [g^2]) \, , \qquad E_i: ([g^1], [g^0], [g^1]) \, , \qquad N_i: ([g^0], [g^1], [g^2]) \, , \nonumber \\
H_u &: [g^1] \, , \qquad \qquad \qquad ~ H_d: [g^1] \, , \label{eq:Z5-1assign}
\end{align}
for $\tilde{\mathbb{Z}}_5^{(1)}$, and
\begin{align}
Q_i &: ([g^2], [g^1], [g^1]) \, , \qquad U_i: ([g^1], [g^2], [g^0]) \, , \qquad D_i: ([g^2], [g^2], [g^0]) \, , \nonumber \\
L_i &: ([g^2], [g^1], [g^1]) \, , \qquad E_i: ([g^1], [g^2], [g^0]) \, , \qquad N_i: ([g^0], [g^2], [g^2]) \, , \nonumber \\
H_u &: [g^1] \, , \qquad \qquad \qquad ~ H_d: [g^1] \, , \label{eq:Z5-2assign}
\end{align}
for $\tilde{\mathbb{Z}}_5^{(2)}$. 
Here, we focus on a specific assignment to realize the realistic flavor patterns, but one can consider other choices, as summarized in Appendix~\ref{app}.\footnote{For illustrative purposes, we focus on $\tilde{\mathbb{Z}}_5^{(1)} \times \tilde{\mathbb{Z}}_5^{(2)}$ to realize novel Yukawa textures~\eqref{eq:Yftexture} that cannot be realized by conventional group-based symmetries. 
Other selection rules will lead to different Yukawa textures as discussed in e.g., Ref.~\cite{Kobayashi:2025znw}, but comprehensive studies will be left for future work.} 
The superpotential of matter fields is written as
\begin{align}
W = Y_u Q H_u U + Y_d Q H_d D + Y_e L H_d E + Y_\nu L H_u N + M_N N N + \mu H_u H_d \, ,
\label{eq:WMSSM}
\end{align}
where the $\mu$-term is allowed under the non-invertible selection rule, and one can obtain the constrained Yukawa textures, as will be explicitly shown later. 

We assume that the non-invertible selection rule is not violated in the SUSY breaking sector. 
Then, the structures of the following soft SUSY breaking terms,
\begin{align}
\mathcal{L}_{\rm soft} \supset - \sum_f M_f^2 \tilde{f}^{\dagger} \tilde{f} - \Bigl( A_u \tilde{Q} H_u \tilde{u} + A_d \tilde{Q} H_d \tilde{d} + A_e \tilde{L} H_d \tilde{e} + {\rm c.c.} \Bigr) \, ,
\label{eq:Lagsoft}
\end{align}
are also constrained by our non-invertible selection rule. 
Here $\tilde{f}$ for $f = Q, u, d, L, e$ represents a scalar partner of the SM fermion, $M_f^2$ are the $3 \times 3$ soft mass-squared matrices for $(Q, u, d, L, e)$ types, and $A_{u, d, e}$ are $3 \times 3$ scalar trilinear couplings, called $A$-terms. 
In the present paper, we focus only on how our non-invertible selection rule suppresses SUSY contributions to FCNC processes, and hence we implicitly assume that all (potentially complex) parameters except for the Yukawa matrices are real. 
We will revisit this issue in section~\ref{conclusion}.

\subsection{Yukawa textures and RG effects}
\label{sec:Yukawatexture}

Using the $\tilde{\mathbb{Z}}_5^{(1)} \times \tilde{\mathbb{Z}}_5^{(2)}$ assignment in Eqs.~\eqref{eq:Z5-1assign} and \eqref{eq:Z5-2assign}, we obtain the following textures for Yukawa couplings:
\begin{align}
Y_u = \begin{pmatrix}
* & 0 & 0 \\
0 & * & 0 \\
0 & 0 & *
\end{pmatrix} \, , ~~ 
Y_d = \begin{pmatrix}
0 & * & 0 \\
* & * & * \\
0 & * & *
\end{pmatrix} \, , ~~ 
Y_e = \begin{pmatrix}
* & 0 & 0 \\
0 & * & 0 \\
0 & 0 & *
\end{pmatrix} \, , ~~ 
Y_{\nu} = \begin{pmatrix}
0 & * & 0 \\
* & 0 & * \\
0 & * & *
\end{pmatrix} \, ,
\label{eq:Yftexture}
\end{align}
where $*$ denotes a nonzero component. 
It was known that these textures, including the nearest neighbor interaction (NNI) type~\cite{Branco:1988iq,Branco:1994jx,Branco:1999nb} and its extension, cannot be realized by a group-based symmetry, and it is a characteristic feature from $\tilde{\mathbb{Z}}_5$ symmetry. 
Let us comment on why it is difficult to realize the above textures using standard group-based symmetries.
For illustrative purposes, we focus on the upper-left $2 \times 2$ submatrix of $(Y_d)_{ij}$.
When we consider a $U(1)$ symmetric model where each matter superfield $\Phi$ carries a $U(1)$ charge $q^{\Phi}_i$, the structure of $(Y_d)_{ij}$ for $i,j=1,2$ can be realized only if the following charge conservation conditions are satisfied:
\begin{align}
q^{Q}_1 + q^{D}_1 + q^{H_d} \neq 0 \, , \quad q^{Q}_2 + q^{D}_1 + q^{H_d} = 0 \, , \quad q^{Q}_1 + q^{D}_2 + q^{H_d} = 0 \, , \quad q^{Q}_2 + q^{D}_2 + q^{H_d} = 0 \, .
\end{align}
However, there is no solution for these charges that can realize the NNI texture and its extension within $U(1)$ symmetric models. This conclusion also applies to non-Abelian symmetries. 
These textures are applied at a high scale, e.g., the scale of gauge coupling unification (we call it as the unification scale in the following discussion) or the string scale. 
We here assume that the model has the above textures at the unification scale, $M_{\rm U} \simeq 2 \times 10^{16} \, {\rm GeV}$.\footnote{Our $\tilde{\mathbb{Z}}_5^{(1)} \times \tilde{\mathbb{Z}}_5^{(2)}$ assignment for matter fields is apparently not consistent with a grand unified theory. 
We will comment on this issue in section~\ref{conclusion}.}

At the electroweak scale, the Yukawa textures of Eq.~\eqref{eq:Yftexture} could be destroyed by renormalization group (RG) effects. 
As pointed out in Ref.~\cite{Xing:2015sva}, however, this kind of texture zero structures is stable against RG effects from the unification scale to the electroweak scale. 
We then expect that our textures in Eq.~\eqref{eq:Yftexture} are also stable. 
To see this explicitly, we consider the one-loop RG running of Yukawa matrices in the MSSM. 
We neglect the effect of $Y_{\nu}$, because it significantly depends on the right-handed neutrino mass scale. 
If we consider the simplest type-I seesaw mechanism~\cite{Minkowski:1977sc,Yanagida:1979as,Yanagida:1980xy,Mohapatra:1979ia,Gell-Mann:1979vob,Schechter:1980gr} to obtain tiny neutrino masses, $M_{N_R} \sim 10^{13} \, {\rm GeV}$ is required, and hence, the effects of the other Yukawa couplings are more relevant for RG evolution than that of $Y_{\nu}$.\footnote{The RG evolution  including $Y_{\nu}$ and the associated soft SUSY breaking terms, $A_{\nu}$ and $M_{\nu}^2$, can be found in Refs.~\cite{Casas:2000pa,Antusch:2002rr,Antusch:2005gp,Antusch:2015nwi}.}
To find Yukawa matrices at the unification scale from the low scale inputs, we utilize the SM input parameters at the soft mass scale \cite{Antusch:2025fpm},\footnote{We use the results of Ref.~\cite{Antusch:2025fpm} for Yukawa couplings in the $\overline{{\rm DR}}$ renormalization scheme, since the MSSM RGEs in Ref.~\cite{Martin:1993zk} are given in that scheme for running parameters.} which are obtained from the experimental results summarized in Particle Data Group 2024~\cite{ParticleDataGroup:2024cfk}. 
Here, we take the basis where Yukawa matrices $Y_u$ and $Y_e$ are diagonal at the soft mass scale, $Y_{u, e} (M_{\rm SUSY}) = {\rm diag}(y_{u, e}, y_{c, \mu}, y_{t, \tau})$, while $Y_d$ has off-diagonal elements originated from the Cabibbo-Kobayashi-Maskawa (CKM) matrix, $Y_d (M_{\rm SUSY}) = V_{\rm CKM}^{\dagger} {\rm diag}(y_d, y_s, y_b)$, with $y_f \, (f = u, c, t, d, s, b, e, \mu, \tau)$ being eigenvalues of Yukawa matrices. 
Then, we convert the SM Yukawa matrices to the MSSM ones through $Y_u^{\rm MSSM} = Y_u / \sin \beta$ and $Y_{d, e}^{\rm MSSM} = Y_{d, e} / \cos \beta$, using $\tan \beta \equiv v_u / v_d$ which is the ratio of two vacuum expectation values (VEVs) of neutral Higgs scalars $H_{u, d}^0$. 
In our notation, these VEVs satisfy the relation $v_u^2 + v_d^2 \simeq (174 \, {\rm GeV})^2$. 
Once Yukawa matrices at the unification scale $M_{\rm U}$ are obtained by using the MSSM RG equations (RGEs)~\cite{Martin:1993zk}, we can transform the Yukawa matrices to match with the desired Yukawa textures given in Eq.~\eqref{eq:Yftexture}. 
It is easy to obtain the $Y_u$ texture by diagonalizing $Y_u (M_{\rm U})$, while the texture of $Y_d$ at the unification scale is obtained by a unitary transformation for the right-handed down-type quarks so that $(Y_d)_{11, 13, 31} = 0$. 
On the other hand, $Y_e$ is diagonal at any scale, because there is no source to generate off-diagonal elements of $Y_e$ in the current setup. 
Therefore, the $Y_e$ texture is totally stable against the MSSM RGEs. 
Note that this feature can be changed when we include RG effects of non-diagonal form of $Y_{\nu}$ in Eq.~\eqref{eq:Yftexture}, which are expected to be small and depend on an explicit model for the neutrino sector. 

Taking account of all rephasing degrees of freedom, we obtain the parameters at the unification scale as
\begin{align}
g_1 (M_{\rm U}) &\simeq 0.699 \, , \quad g_2 (M_{\rm U}) \simeq 0.686 \, , \quad g_3 (M_{\rm U}) \simeq 0.688 \, , \\[1.5ex]
Y_u^{\rm MSSM} (M_{\rm U}) &\simeq {\rm diag} \left( 2.89 \times 10^{-6}, 1.46 \times 10^{-3}, 0.527 \right) \, , \label{eq:YuGUT3TeV} \\[0.5ex]
Y_d^{\rm MSSM} (M_{\rm U}) &\simeq \begin{pmatrix}
0 & 0.0256 & 0 \\
0.0109 & 0.118 e^{1.09 i} & 0.0199 \\
0 & 2.53 & 1.27
\end{pmatrix} \times 10^{-2} \, , \label{eq:YdGUT3TeV} \\[0.5ex]
Y_e^{\rm MSSM} (M_{\rm U}) &\simeq {\rm diag} \left( 1.02 \times 10^{-5}, 2.15 \times 10^{-3}, 3.66 \times 10^{-2} \right) \, , \label{eq:YeGUT3TeV}
\end{align}
where $g_1 = \sqrt{5 / 3} g'$ is the normalized $U(1)_{Y}$ gauge coupling, and we choose $\tan \beta = 5$ and $M_{\rm SUSY} = 3 \, {\rm TeV}$. 
One phase in $Y_d^{\rm MSSM}$ cannot be removed, which corresponds to the phase in the CKM matrix. 
The number of independent parameters for Yukawa matrices is $13$. 
We use the above values as input parameters at the unification scale and find the following Yukawa textures at the soft mass scale $M_{\rm SUSY} = 3 \, {\rm TeV}$:
\begin{align}
|Y_u (M_{\rm SUSY})| &\simeq \begin{pmatrix}
5.78 \times 10^{-6} & 4.55 \times 10^{-10} & 2.59 \times 10^{-6} \\
9.01 \times 10^{-13} & 2.92 \times 10^{-3} & 1.25 \times 10^{-5} \\
1.46 \times 10^{-11} & 3.55 \times 10^{-8} & 0.816
\end{pmatrix} \, , \label{eq:AbsYu3TeV} \\
|Y_d (M_{\rm SUSY})| &\simeq \begin{pmatrix}
2.58 \times 10^{-11} & 1.29 \times 10^{-4} & 5.75 \times 10^{-8} \\
5.51 \times 10^{-5} & 5.94 \times 10^{-4} & 1.00 \times 10^{-4} \\
2.40 \times 10^{-9} & 1.17 \times 10^{-2} & 5.88 \times 10^{-3}
\end{pmatrix} \, , \label{eq:AbsYd3TeV} \\[0.5ex]
|Y_e (M_{\rm SUSY})| &\simeq {\rm diag} \left( 2.840 \times 10^{-6}, 5.982 \times 10^{-4}, 1.016 \times 10^{-2} \right) \, . \label{eq:AbsYe3TeV}
\end{align}
Here, we have shown the SM Yukawa matrices by using $Y_u = Y_u^{\rm MSSM} \sin \beta$ and $Y_{d, e} = Y_{d, e}^{\rm MSSM} \cos \beta$, and the resulting eigenvalues of each Yukawa matrix and CKM parameters are consistent with those in Ref.~\cite{Antusch:2025fpm}. 
Although all elements in $Y_{u, d}$ become nonzero due to the RG effects, their structures are similar to those at the unification scale. In fact, we have found that mixing angles in diagonalizing matrices for $Y_u$ are sufficiently small and up to $\mathcal{O} (10^{-5})$, and $(Y_d)_{11, 13, 31}$ elements are much smaller than the other elements. 
Hence, the flavor structure of quarks and leptons remains fixed by the non-invertible selection rule, even when loop effects violate them.

\subsection{Soft SUSY breaking}
\label{sec:SUSYpara}

Our $\tilde{\mathbb{Z}}_5^{(1)} \times \tilde{\mathbb{Z}}_5^{(2)}$ assignments are also applied to partner particles, sfermions, under our assumption explained in section~\ref{sec:setup}. 
Therefore, all matrices of the soft SUSY breaking terms in Eq.~\eqref{eq:Lagsoft} have structures constrained by our selection rule. 
It is clear that the $A$-terms have the same textures as the corresponding Yukawa matrices in Eq.~\eqref{eq:Yftexture}. 
On the other hand, all soft mass-squared matrices for sfermions are diagonal, as one can easily check from our assignments in Eqs.~\eqref{eq:Z5-1assign} and \eqref{eq:Z5-2assign}. 
These diagonal forms of soft mass-squared matrices are qualitatively different from those in standard approaches, such as the Froggatt-Nielsen mechanism~\cite{Froggatt:1978nt}.\footnote{In models with ordinary symmetries, a spurion field $S$ that breaks the horizontal symmetry typically induces off-diagonal entries in the soft mass-squared matrices, through terms in the K\"ahler potential like $K \supset \frac{\langle S \rangle^{\dagger} \langle S \rangle}{\Lambda^2} \tilde{f}^{\dagger}_i \tilde{f}_j$ (even for $i \neq j$) with a high energy scale $\Lambda$. 
The spurion VEVs $\langle S \rangle \sim \lambda \Lambda$, required to explain the Yukawa hierarchy via a small factor $\lambda \sim \mathcal{O} (0.1)$, simultaneously generate flavor-violating soft terms. 
This feature is generic to models with flavor symmetries that govern the Yukawa textures. 
In contrast, the selection rule in our scenario is applied directly to the correlation functions of the matter fields, and the specific charge assignments forbid any off-diagonal combinations in the K\"ahler potential, thereby leading to vanishing FCNC contributions at the symmetry level.} 
We emphasize that for these soft SUSY breaking parameters, the non-invertible selection rule can only constrain zero or nonzero for each element but cannot control the size of a nonzero entry: for example, although we can obtain diagonal soft mass-squared matrices, they are in general not proportional to the unit matrix only with our non-invertible selection rule. 

In our analyses, we assume that $A$-terms have similar hierarchies to the corresponding Yukawa matrices, and each nonzero element has random $\mathcal{O} (1)$ ambiguity. 
Similarly, the soft mass-squared matrices have individual $\mathcal{O} (1)$ ambiguity for each diagonal element. 
In order to avoid tachyonic squarks or sleptons, all diagonal elements in their soft mass-squared matrices must be positive, while the $A$-terms can be either positive or negative. 
Therefore, our $A$-terms and soft mass-squared matrices at the unification scale are
\begin{align}
(A_f (M_{\rm U}))_{ij} &= \sigma_A M_{\rm SUSY} \mathcal{O} (1) (Y_f (M_{\rm U}))_{ij} &&\text{for $f = u, d, e$} \, , \label{eq:GUTAterm} \\[1ex]
M_f^2 (M_{\rm U}) &= M_{\rm SUSY}^2 {\rm diag} \left( \mathcal{O} (1), \mathcal{O} (1), \mathcal{O} (1) \right) &&\text{for $f = Q, u, d, L, e$} \, , \label{eq:GUTMsf2}
\end{align}
where the range of $\mathcal{O} (1)$ is $(0,1]$, and $\sigma_A = +1$ or $-1$ denotes the random sign for each element of the $A$-terms. 
For simplicity, we set common mass scales for the $A$-terms and the soft mass-squared matrices as $M_{\rm SUSY}$ and $M_{\rm SUSY}^2$, respectively. 
It is clear that the above textures for $A_f$ and $M_f^2$ do not remain the same at the soft mass scale due to their RG running, although its effects are expected to be mild, as in the case of RG effects on the Yukawa couplings. 

The remaining soft SUSY breaking parameters relevant for the MSSM RGEs are three gaugino masses and mass-squared parameters for $H_{u, d}$,
\begin{align}
\mathcal{L}_{\rm soft} \supset - \frac{1}{2} \sum_{a = 1}^3 M_a \lambda_a \lambda_a - m_{H_u}^2 H_u^{\dagger} H_u - m_{H_d}^2 H_d^{\dagger} H_d  \, . \label{eq:Lsoftadd}
\end{align}
Here, $\lambda_a \, (a =1,2,3)$ denote gauginos, and $H_{u, d}$ are the scalar components of two Higgs doublet chiral superfields. 
For these mass parameters, we assume that $M_{1, 2, 3} = M_{\rm SUSY}$, $m_{H_u}^2 = m_{H_d}^2 = M_{\rm SUSY}^2$ at the unification scale. 
Note that $m_{H_u}^2$ is driven to be negative at the soft mass scale by RG effects of a large top Yukawa coupling as well as a large $A$-term, so that the correct electroweak symmetry breaking is realized. 

Let us comment on the supersymmetric $\mu$-term. 
Our assignments of $\tilde{\mathbb{Z}}_5^{(1)} \times \tilde{\mathbb{Z}}_5^{(2)}$ allow a nonzero $\mu$-term in the superpotential, Eq.~\eqref{eq:WMSSM}, and $\mu \sim M_{\rm SUSY}$ is required for the correct electroweak symmetry breaking. 
Since the $\mu$-term does not affect the one-loop MSSM RGEs for the parameters ($g_i$, $Y_f$, $M_i$, $A_f$, $M_f^2$, $m_{H_{u, d}}^2$), we determine $\mu$ by the following tree-level expression given in terms of $m_{H_{u, d}}^2$ at the soft mass scale and $\tan \beta$:
\begin{align}
\mu = \sqrt{\frac{1}{2} \left[ \tan (2 \beta) \left( m_{H_u}^2 \tan \beta - \frac{m_{H_d}^2}{\tan \beta} \right) - M_Z^2 \right]} \, ,
\label{eq:mutree}
\end{align}
where $M_Z$ is the $Z$ boson mass. 
Note that for $m_{H_u}^2 < 0$ and $m_{H_d}^2 > 0$, the inside of the square root is positive as long as $|m_{H_u}^2|$ and $m_{H_d}^2$ are much larger than $M_Z^2$, because of $\tan (2 \beta) < 0$ for $\tan \beta > 1$.

\subsection{Mass insertion parameters}
\label{sec:MIpara}

In order to discuss SUSY contributions to FCNC processes in our setup, it is convenient to utilize the mass insertion (MI) parameters which are defined by the $6 \times 6$ sfermion mass-squared matrices. 
Each matrix is calculated as
\begin{align}
\mathcal{M}_u^2 &= \begin{pmatrix}
\hat{M}_Q^2 + v_u^2 \hat{Y}_u \hat{Y}_u + \mathcal{D}_{\tilde{u}_L} & v_u \hat{A}_u^* - \mu v_d \hat{Y}_u \\
v_u \hat{A}_u^T - \mu^* v_d \hat{Y}_u & \hat{M}_u^2 + v_u^2 \hat{Y}_u \hat{Y}_u + \mathcal{D}_{\tilde{u}_R}
\end{pmatrix} \equiv \begin{pmatrix}
\Delta_{LL}^u & \Delta_{LR}^u \\
(\Delta_{LR}^u)^{\dagger} & \Delta_{RR}^u
\end{pmatrix} \, , \label{eq:msu6x6} \\
\mathcal{M}_d^2 &= \begin{pmatrix}
V_{\rm CKM}^{\dagger} \hat{M}_Q^2 V_{\rm CKM} + v_d^2 \hat{Y}_d \hat{Y}_d + \mathcal{D}_{\tilde{d}_L} & - v_d \hat{A}_d^* + \mu v_u \hat{Y}_d \\
- v_d \hat{A}_d^T + \mu^* v_u \hat{Y}_d & \hat{M}_d^2 + v_d^2 \hat{Y}_d \hat{Y}_d + \mathcal{D}_{\tilde{d}_R}
\end{pmatrix} \equiv \begin{pmatrix}
\Delta_{LL}^d & \Delta_{LR}^d \\
(\Delta_{LR}^d)^{\dagger} & \Delta_{RR}^d
\end{pmatrix} \, , \label{eq:msd6x6} \\
\mathcal{M}_e^2 &= \begin{pmatrix}
\hat{M}_L^2 + v_d^2 \hat{Y}_e \hat{Y}_e + \mathcal{D}_{\tilde{e}_L} & - v_d \hat{A}_e^* + \mu v_u \hat{Y}_e \\
- v_d \hat{A}_e^T + \mu^* v_u \hat{Y}_e & \hat{M}_e^2 + v_d^2 \hat{Y}_e \hat{Y}_e + \mathcal{D}_{\tilde{e}_R}
\end{pmatrix} \equiv \begin{pmatrix}
\Delta_{LL}^e & \Delta_{LR}^e \\
(\Delta_{LR}^e)^{\dagger} & \Delta_{RR}^e
\end{pmatrix} \, , \label{eq:mse6x6}
\end{align}
where each block is a $3 \times 3$ matrix, and we take a basis in which all the SM Yukawa matrices are real and diagonal, $\hat{Y}_f = L_f^{\dagger} Y_f R_f$, and the same rotations are applied to the corresponding sfermions.\footnote{Sometimes this basis is called ``super-CKM basis".}
Then, $\hat{M}_f^2$ and $\hat{A}_f$ are defined as
\begin{align}
\hat{M}_Q^2 \equiv L_u^{\dagger} M_Q^2 L_u \, , \quad \hat{M}_u^2 \equiv R_u^{\dagger} M_u^2 R_u \, , \quad \hat{A}_u \equiv L_u^{\dagger} A_u R_u \, .
\end{align}
The others are similarly defined. 
In the above matrices, $\mathcal{D}_{\phi}$ in each diagonal block is a $D$-term contribution which can be calculated for each sfermion, $\phi = \tilde{u}_L, \tilde{u}_R, \tilde{d}_L, \tilde{d}_R, \tilde{e}_L, \tilde{e}_R$, as
\begin{align}
\mathcal{D}_{\phi} = \left[ \frac{g_2^2 + (3/5) g_1^2}{2} \left( T_3 (\phi) - Q (\phi) \sin^2 \theta_W \right) \cos (2 \beta) v_H^2 \right] \times \mathbbm{1}_3 \, ,
\label{eq:Dterm}
\end{align}
where $T_3 (\phi)$ and $Q (\phi)$ denote the third component of the weak isospin and the electric charge of $\phi$, respectively, $\sin^2 \theta_W = 3 g_1^2 / (5 g_2^2 + 3 g_1^2)$, $v_H^2 \equiv v_u^2 + v_d^2$, and $\mathbbm{1}_3$ is the $3 \times 3$ unit matrix. 
Although these contributions are flavor-blind and smaller than the soft mass-squared matrices, we include them in our analyses. 
Then, the MI parameters can be defined by
\begin{align}
(\delta_{XY}^f)_{ij} \equiv \frac{(\Delta_{XY}^f)_{ij}}{m_{\tilde{f}}^2} \, ,
\end{align}
for $f = u, d, e$ and $XY = LL, RR, LR$. 
Here, $m_{\tilde{f}}^2$ is an average of eigenvalues of $\mathcal{M}_f^2$, which can be considered as a typical mass scale for each sector. 
It is expected that off-diagonal elements of $\mathcal{M}_f^2$ are smaller than diagonal ones, and $m_{\tilde{f}}^2$ is almost determined by the diagonal elements of $\mathcal{M}_f^2$. 
This leads to $(\delta_{LL, RR}^f)_{ii} \simeq 1$, $(\delta_{LL, RR}^f)_{ij} \ll 1$ (for $i \neq j$) and $(\delta_{LR}^f)_{ij} \ll 1$. 
As long as these relations are satisfied, the estimation for SUSY contributions to FCNC processes via the MI parameters is in good agreement with a full calculation which can be performed by diagonalizing $\mathcal{M}_f^2$ and using their eigenvalues. 

Using the inputs at the unification scale in Eqs.~\eqref{eq:GUTAterm} and \eqref{eq:GUTMsf2}, we generate 10,000 samples with random $\mathcal{O} (1)$ factors as well as random signs for soft breaking parameters, and obtain the following mean values of the MI parameters and their 95\% confidence interval (CI) shown in parentheses for the quark sector:
\begin{align}
|\delta^u_{LL}| &= \begin{pmatrix}
1.2(1.0,1.3) & 7.1(5.5,9.5) \times 10^{-7} & 1.3(1.0,1.8) \times 10^{-5} \\
7.1(5.5,9.5) \times 10^{-7} & 1.2(1.0,1.3) & 6.0(4.6,8.2) \times 10^{-5} \\
1.3(1.0,1.8) \times 10^{-5} & 6.0(4.6,8.2) \times 10^{-5} & 0.94(0.82,1.07)
\end{pmatrix} \, , \label{eq:deluLLatSUSY} \\
|\delta^u_{RR}| &= \begin{pmatrix}
1.1(0.9,1.2) & 5.9(0.2,14.8) \times 10^{-11} & 6.0(0.2,16.2) \times 10^{-12} \\
5.9(0.2,14.8) \times 10^{-11} & 1.1(0.9,1.2) & 8.1(5.9,10.5) \times 10^{-8} \\
6.0(0.2,16.2) \times 10^{-12} & 8.1(5.9,10.5) \times 10^{-8} & 0.63(0.50,0.76)
\end{pmatrix} \, , \label{eq:deluRRatSUSY} \\
|\delta^u_{LR}| &= \begin{pmatrix}
2.1(1.2,3.1) \times 10^{-7} & 4.8(2.7,7.0) \times 10^{-11} & 1.1(0.3,1.9) \times 10^{-7} \\
9.7(5.5,13.7) \times 10^{-14} & 1.1(0.6,1.6) \times 10^{-4} & 1.1(0.7,1.5) \times 10^{-6} \\
7.8(0.5,16.9) \times 10^{-13} & 3.0(1.9,4.2) \times 10^{-9} & 2.4(1.9,3.0) \times 10^{-2}
\end{pmatrix} \, , \label{eq:deluLRatSUSY} \\
|\delta^d_{LL}| &= \begin{pmatrix}
1.1(1.0,1.2) & 1.97(0.08,4.94) \times 10^{-2} & 1.2(0.3,2.2) \times 10^{-3} \\
1.97(0.08,4.94) \times 10^{-2} & 1.1(1.0,1.2) & 8.7(0.7,17.0) \times 10^{-3} \\
1.2(0.3,2.2) \times 10^{-3} & 8.7(0.7,17.0) \times 10^{-3} & 0.89(0.77,1.01)
\end{pmatrix} \, , \label{eq:deldLLatSUSY} \\
|\delta^d_{RR}| &= \begin{pmatrix}
0.98(0.86,1.10) & 1.71(0.07,4.25) \times 10^{-2} & 7.5(0.3,19.2) \times 10^{-3} \\
1.71(0.07,4.25) \times 10^{-2} & 0.98(0.89,1.08) & 3.4(0.1,8.8) \times 10^{-2} \\
7.5(0.3,19.2) \times 10^{-3} & 3.4(0.1,8.8) \times 10^{-2} & 0.98(0.88,1.08)
\end{pmatrix} \, , \label{eq:deldRRatSUSY} \\
|\delta^d_{LR}| &= \begin{pmatrix}
1.2(0.8,1.6) \times 10^{-6} & 6.8(5.1,8.4) \times 10^{-6} & 8.4(5.7,11.0) \times 10^{-6} \\
6.5(4.2,8.9) \times 10^{-6} & 3.5(2.4,4.7) \times 10^{-5} & 5.0(3.9,6.2) \times 10^{-5} \\
1.8(1.2,2.3) \times 10^{-4} & 8.2(5.7,10.7) \times 10^{-4} & 9.0(7.0,11.0) \times 10^{-4}
\end{pmatrix} \, , \label{eq:deldLRatSUSY}
\end{align}
for the case of $M_{\rm SUSY} = 3 \, {\rm TeV}$ with $\tan \beta = 5$ and $\mu > 0$. 
Note that the choice of the sign for $\mu$ affects $\delta_{LR}^f$, but the resulting values of elements are similar to the above ones. 
The off-diagonal elements for the up-type MI parameters are sufficiently smaller than the diagonal elements, while those for the down-type MI parameters have non-trivial values. 
It is notable that these structures are representative predictions of our model, originating from the Yukawa textures and diagonal soft mass-squared matrices.\footnote{In models with the alignment mechanism based on horizontal symmetries~\cite{Leurer:1992wg,Nir:1993mx,Leurer:1993gy,Ibanez:1994ig,Nakai:2021mha}, the off-diagonal elements of $\delta^u_{LL, RR}$ are generally larger than those predicted in Eqs.~\eqref{eq:deluLLatSUSY} and \eqref{eq:deluRRatSUSY}, whereas those of $\delta^d_{LL, RR}$ are smaller than our predictions in Eqs.~\eqref{eq:deldLLatSUSY} and \eqref{eq:deldRRatSUSY}. 
As will be discussed later, these sizes are directly reflected in the predictions of the SUSY FCNC processes: our model will have smaller predictions of $D^0$--$\overline{D^0}$ mixing and larger ones of $K^0$--$\overline{K^0}$ mixing, where the latter ones are close to the current experimental bounds. 
It is worth emphasizing that, unlike models based on horizontal symmetries, the SUSY FCNC predictions in our model are essentially fixed up to $\mathcal{O}(1)$ uncertainties.}  
Therefore, flavor observables related to down-type (s)quarks will be interesting in our model. 
Note that $(\delta_{LL, RR}^u)_{33}$ and $(\delta_{LL}^d)_{33}$ tend to be smaller than the other diagonal elements, because of large negative RG effects proportional to the top Yukawa coupling. 
Nevertheless, they are still larger than the off-diagonal elements, and hence, the MI approximation will work well for the estimation of SUSY contributions to FCNC processes. 

For the lepton sector, there are no sources to induce off-diagonal elements for $A_e$ as well as $M_{L, e}^2$ in the MSSM RG running, and hence, only diagonal elements of $A_e$ and $M_{L, e}^2$ are modified. 
For completeness, we present our results of MI parameters for the lepton sector at the soft mass scale as follows:
\begin{align}
|\delta^e_{LL}| &= {\rm diag} \Bigl( 1.20 (0.63, 1.87), 1.20 (0.63, 1.86), 1.20 (0.62, 1.85) \Bigr) \, , \label{eq:deltaeLL} \\[0.5ex]
|\delta^e_{RR}| &= {\rm diag} \Bigl( 0.80 (0.22, 1.42), 0.81 (0.22, 1.41), 0.80 (0.21, 1.42) \Bigr) \, , \label{eq:deltaeRR} \\[0.5ex]
|\delta^e_{LR}| &= {\rm diag} \Bigl( 1.11 (0.67, 1.69) \times 10^{-6}, 2.34 (1.42, 3.58) \times 10^{-4}, 3.97 (2.42, 6.09) \times 10^{-3} \Bigr) \, , \label{eq:deltaeLR}
\end{align}
with the same notation as in Eqs.~\eqref{eq:deluLLatSUSY}-\eqref{eq:deldLRatSUSY}. 
Although this conclusion is affected by introducing $Y_{\nu}$, $A_{\nu}$ and $M_{\nu}^2$, these effects are expected to be small and negligible in our analyses.

\section{Contributions to flavor observables}
\label{FCNC}

With MI parameters found in the previous section, we now discuss SUSY contributions to various flavor observables in our setup. 
Our estimation of each observable via the MI parameters is checked with the result obtained by the public code \verb|susy_flavor_v2.54|~\cite{Rosiek:2010ug,Crivellin:2012jv,Rosiek:2014sia}. 
Since no CP-violating phase in the SUSY parameters is introduced, we mostly focus on CP-conserving processes, e.g., meson mixing parameters and branching ratios for $\Delta F = 1$ FCNC processes. 
However, in the squark sector, we have a CP violation source originated from the phase in the CKM matrix, as we can explicitly see in Eqs.~\eqref{eq:msu6x6} and \eqref{eq:msd6x6}. 
Hence, we also check SUSY contributions to observables related to CP violation in the quark sector, which are the neutron electric dipole moment (EDM) and the CP-violating parameter $\varepsilon_K$ in $K^0$ system. 
In our estimation, we take the sfermion mass scale $m_{\tilde{f}} \sim M_{\rm SUSY} = 3 \, {\rm TeV}$ as a reference value.

\subsection{Lepton observables}
\label{sec:FCNC-lepton}

Since all parameters in the lepton sector are diagonal, all MI parameters at the soft mass scale are also diagonal, as shown in Eqs.~\eqref{eq:deltaeLL}-\eqref{eq:deltaeLR}. 
Therefore, SUSY contributions to lepton flavor violating (LFV) processes, like $\mu \to e \gamma$ and $\mu \to 3e$, vanish in our setup. 
As mentioned in section~\ref{model}, when we introduce neutrino and sneutrino parameters in our analysis, off-diagonal elements of the MI parameters are induced, so that the model is potentially constrained by the LFV observables. 
Using the estimated values for slepton masses, $\mu$-term and bino mass as well as the current experimental bound of ${\rm BR} (\mu \to e \gamma) < 3.1 \times 10^{-13}$~\cite{ParticleDataGroup:2024cfk,MEGII:2023ltw}, the off-diagonal elements of slepton MI parameters are constrained as $|\delta^e| \lesssim 10^{-2}$, according to the studies in Refs.~\cite{Nakai:2021mha,Paradisi:2005fk}. 

Let us comment on flavor-conserving observables, anomalous magnetic moments $a_{\ell} \equiv (g-2)_{\ell}$. 
SUSY contributions to $a_{\ell}$ are roughly estimated as~\cite{Moroi:1995yh,Martin:2001st}
\begin{align}
\frac{a_{\ell}^{\rm SUSY}}{m_{\ell}^2} \approx \frac{6.5 \times 10^{-10}}{{\rm GeV}^2} \times \left( \frac{\tan \beta}{5} \right) \left( \frac{3 \, {\rm TeV}}{M_{\rm SUSY}} \right)^2 \, ,
\end{align}
where we assume that all supersymmetric particles have a common mass of $M_{\rm SUSY}$. 
This approximation is in good agreement with that obtained by \verb|susy_flavor_v2.54|, which is $a_e^{\rm SUSY} \simeq 2.1 \times 10^{-16}$, $a_{\mu}^{\rm SUSY} \simeq 8.8 \times 10^{-12}$ and $a_{\tau}^{\rm SUSY} \simeq 2.5 \times 10^{-9}$. 
These values are smaller than the observed values~\cite{Fan:2022eto,Muong-2:2025xyk,ATLAS:2022ryk,CMS:2024qjo}.

\subsection{Up-type quark observables}
\label{sec:FCNC-upquark}

For the up-type squark sector, we have off-diagonal MI parameters, as shown in Eqs.~\eqref{eq:deluLLatSUSY}-\eqref{eq:deluLRatSUSY}. 
However, their values are small, since the off-diagonal elements of $Y_u, A_u, M_Q^2, M_u^2$, which are induced by RG effects, are small enough compared with the diagonal elements. 
Therefore, SUSY contributions to FCNC processes related to the up-type quark sector are expected to be suppressed. 
One of such FCNC processes is the top quark decay into a light up-type quark and the Higgs, $t \to q + h$ with $q = u, c$. 
We have found that our model gives ${\rm BR} (t \to c h) = 1.44 \, (1.35, 1.53) \times 10^{-13}$ and ${\rm BR} (t \to u h) = 1.15 \, (1.08, 1.22) \times 10^{-15}$ for mean and 95\% confidence interval (CI), using \verb|susy_flavor_v2.54|. 
These are about 9 and 11 orders of magnitude smaller than the current experimental limits at 95\% confidence level (CL), ${\rm BR} (t \to c h) < 3.4 \times 10^{-4}$ and ${\rm BR} (t \to u h) < 1.9 \times 10^{-4}$~\cite{CMS:2021hug,ATLAS:2024mih}, respectively. 
Therefore, our model is extremely insensitive to these top quark FCNC processes. 

Other relevant FCNC observables for the up-type quark sector are the meson mixing parameters in the $D^0$--$\overline{D^0}$ oscillation. 
In our analysis, we focus on the mass difference of two eigenvalues of dispersive part of time evolution in $D^0$--$\overline{D^0}$ system,\footnote{For more details, including other observables in neutral $D$ meson system, see the review~\cite{ParticleDataGroup:2024cfk}.} whose definition is shown later in Eq.~\eqref{eq:DelM}. 
Such meson mixing parameters are in general dominated by the gluino contributions, due to the strong coupling as well as no CKM suppression in the quark-squark-gluino coupling. 
However, in the current setup, the MI parameters for the up-type squarks, which play a role to change the flavors in the gluino contributions, are highly suppressed, due to exactly diagonal forms of $Y_u, A_u, M_{Q, u}^2$ at the unification scale. 
The same argument is applied to the neutralino contributions. 
Then, the dominant contributions can be given by the chargino or charged Higgs, although their diagrams have CKM suppression
in each vertex. 
Comparing these two contributions, we have found that the charged Higgs ones dominate over the chargino ones, because the latter has additional suppression from MI parameters for down-type squarks and heavy supersymmetric particles in all internal lines. 
By using \verb|susy_flavor_v2.54|, the total SUSY contribution is estimated as
\begin{align}
\Delta M_D^{\rm SUSY} = -1.83 \, (-1.85, -1.80) \times 10^{-18} \, {\rm GeV} \, ,
\label{eq:DelMD}
\end{align}
which is dominated by the charged Higgs contributions and the sign is opposite to the SM prediction. 
It should be, however, noted that the theoretical uncertainty for calculation of $D^0$--$\overline{D^0}$ is still large, and therefore, the SM prediction is not precisely calculated. 
According to the recent study~\cite{Dulibic:2025emg}, the leading order calculation for the SM prediction results in $\Delta M_D^{\rm SM, \, LO} \sim 10^{-18} \, {\rm GeV}$, while the contributions from higher-dimensional operators (HDOs) give a one order of magnitude larger result, $\Delta M_D^{\rm SM, \, HDOs} \sim 10^{-17} \, {\rm GeV}$.\footnote{See, also Refs.~\cite{Hansen:2012tf,DiCarlo:2025mnm} for direct calculation by using Lattice QCD.}
Our SUSY contribution is one order of magnitude smaller than $\Delta M_D^{\rm SM, \, HDOs}$. 
Note that the current experimental result is $\Delta M_D^{\rm exp} = 6.56 (76) \times 10^{-15} \, {\rm GeV}$~\cite{ParticleDataGroup:2024cfk}, and hence, our setup is anyway not sensitive to the $\Delta M_D$ measurement. 
Note that models with horizontal symmetries discussed in Refs.~\cite{Leurer:1992wg,Nir:1993mx,Leurer:1993gy,Ibanez:1994ig,Nakai:2021mha} predict larger $\Delta M_D$ than Eq.~\eqref{eq:DelMD}, due to large off-diagonal elements of $\delta^u_{LL, RR}$. 
This leads their prediction to be dominated by the gluino contributions, which is totally different situation from our model.

\subsection{Down-type quark observables}
\label{sec:FCNC-downquark}

Unlike the lepton and up-type quark sectors, the down-type quark sector potentially gives sizable contributions to FCNC processes, due to large flavor-changing MI parameters. 
We investigate SUSY contributions to various observables including neutral meson oscillations, neutron EDM and $b \to s \gamma$ in the present setup. 

\subsubsection{Neutral meson oscillations}
\label{sec:DelF=2}

Similar to the $D^0$--$\overline{D^0}$ oscillation, there are three neutral meson oscillations in the down-type quark sector, $K^0$--$\overline{K^0}$, $B_d^0$--$\overline{B_d^0}$ and $B_s^0$--$\overline{B_s^0}$, and each mass difference $\Delta M_M$ for $M = K, B_d, B_s$ gives an important observable to test new physics. 
The mass difference of $M^0$ meson $\Delta M_M$ is given by~\cite{Gabbiani:1996hi,Ciuchini:1998ix}
\begin{align}
\Delta M_M = 2 \, {\rm Re} \langle M^0| \mathcal{H}_{\rm eff}^{\Delta F = 2} | \overline{M^0} \rangle \, ,
\label{eq:DelM}
\end{align}
where $\mathcal{H}_{\rm eff}^{\Delta F = 2}$ is the effective Hamiltonian for the $\Delta F = 2$ process. 
In addition to the mass differences for $K$ and $B_{d, s}$, the CP-violating parameter for the Kaon system $\varepsilon_K$, which is described as
\begin{align}
\varepsilon_K = \frac{1}{\sqrt{2} \Delta M_K} {\rm Im} \langle K^0| \mathcal{H}_{\rm eff}^{\Delta F = 2} | \overline{K^0} \rangle \, ,
\end{align}
is also sensitive to new physics. 
Table~\ref{tablemeson} summarizes the current experimental results as well as the corresponding SM predictions for $\Delta M_M$ and $\varepsilon_K$. 
\begin{table}[t!]
\centering
\begin{tabular}{ccc}
\hline
Observable & Exp. results & SM predictions \\
\hline
$\Delta M_K$ & $3.483(6) \times 10^{-15} \, {\rm GeV}$ & $3.1(1.2) \times 10^{-15} \, {\rm GeV}$ \\
$\Delta M_{B_d}$ & $3.33(1) \times 10^{-13} \, {\rm GeV}$ & $3.52(14) \times 10^{-13} \, {\rm GeV}$ \\
$\Delta M_{B_s}$ & $1.1700(4) \times 10^{-11} \, {\rm GeV}$ & $1.20(4) \times 10^{-11} \, {\rm GeV}$ \\ \hline
\multirow{2}{*}{$|\varepsilon_K|$} & \multirow{2}{*}{$2.228(11) \times 10^{-3}$} & $(1.428 \sim 1.653) \times 10^{-3}$ (excl.) \\
 &  & $(2.017 \sim 2.098) \times 10^{-3}$ (incl.) \\
\hline
\end{tabular}
\caption{Experimental results and SM predictions for the meson mixing parameters, $\Delta M_{K, B_d, B_s}$ and $\varepsilon_K$~\cite{ParticleDataGroup:2024cfk,Albrecht:2024oyn,Mescia:2012fg,Jwa:2025fon}. 
For $\varepsilon_K$, there are several methods to determine the SM prediction, e.g., using exclusive (excl.) and inclusive (incl.) values of $|V_{cb}| \equiv |(V_{\rm CKM})_{23}|$, different input parameter sets and $u-t$ or $c-t$ unitarity for the CKM elements~\cite{Jwa:2025fon}. 
See, also Refs.~\cite{Bailey:2015tba,Bailey:2018feb} for previous studies. }
\label{tablemeson}
\end{table}

For Kaon mixing, as discussed in Ref.~\cite{Gabbiani:1996hi}, we can estimate dominant SUSY contributions as
\begin{align}
&| \Delta M^{\rm SUSY}_K | \simeq 10^{-11} \times \left( \frac{3 \, {\rm TeV}}{M_{\rm SUSY}} \right)^2 \times \left| {\rm Re} \Bigl[ (\delta^d_{LL})_{12} (\delta^d_{RR})_{12} \Bigr] \right| \, {\rm GeV} \, , \\[1ex]
&| \varepsilon_K^{\rm SUSY} | \simeq 10^4 \times \left( \frac{3 \, {\rm TeV}}{M_{\rm SUSY}} \right)^2 \times \left| {\rm Im} \Bigl[ (\delta^d_{LL})_{12} (\delta^d_{RR})_{12} \Bigr] \right| \, .
\end{align}
Taking the mean value of combination $(\delta^d_{LL})_{12} (\delta^d_{RR})_{12}$, we have $\left| {\rm Re} \Bigl[ (\delta^d_{LL})_{12} (\delta^d_{RR})_{12} \Bigr] \right| \simeq 10^{-4}$ and $\left| {\rm Im} \Bigl[ (\delta^d_{LL})_{12} (\delta^d_{RR})_{12} \Bigr] \right| \simeq 10^{-7}$, leading to $| \Delta M^{\rm SUSY}_K | \simeq 10^{-15} \, {\rm GeV}$ and $| \varepsilon_K^{\rm SUSY} | \simeq 10^{-3}$, which are comparable to the experimental data shown in Table~\ref{tablemeson}. 
\begin{figure}[!t]
\centering
\includegraphics[width=0.45\textwidth]{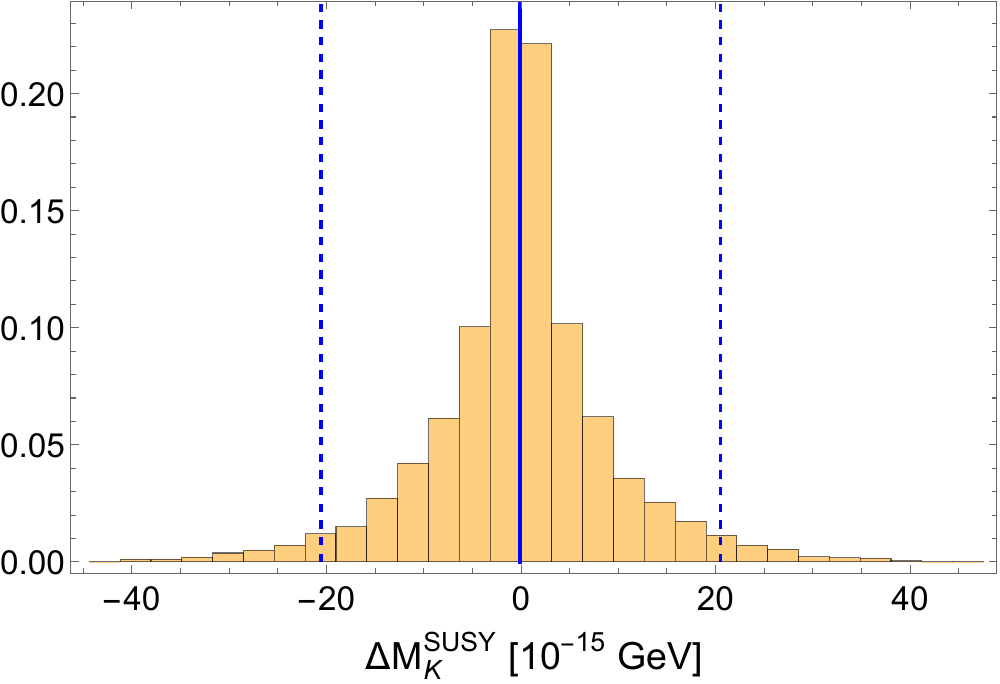} ~~ \includegraphics[width=0.45\textwidth]{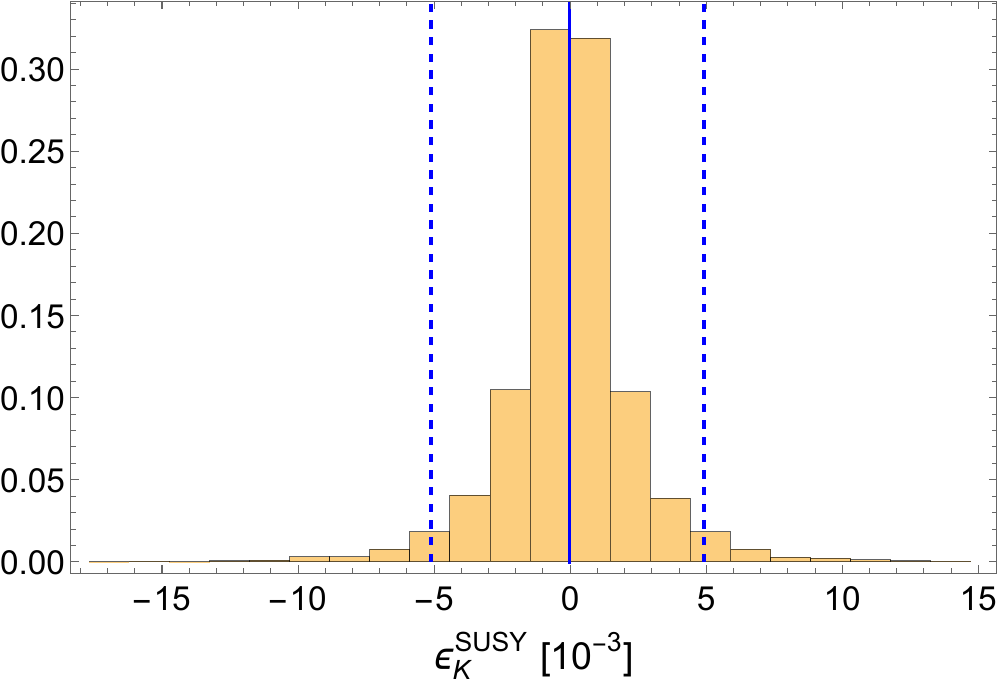}
\caption{Distributions for $\Delta M_K^{\rm SUSY}$ and $\varepsilon_K^{\rm SUSY}$ in our model. 
The sold blue line corresponds to the mean value, and the 95\% CI is between two dashed blue lines. }
\label{fig:KKbar}
\end{figure}
Fig.~\ref{fig:KKbar} shows distributions for $\Delta M_K^{\rm SUSY}$ and $\varepsilon_K^{\rm SUSY}$ generated from the charged Higgs, gluino, neutralino and chargino contributions, subtracting the SM values calculated by \verb|susy_flavor_v2.54|. 
Here, the solid blue line represents the mean value, and dashed ones are 95\% CI (we use the same color codes in the following discussion). 
The mean and 95\% CI values for $\Delta M_K^{\rm SUSY}$ and $\varepsilon_K^{\rm SUSY}$ are obtained as
\begin{align}
&\Delta M_K^{\rm SUSY} = -0.1 \, (-20.6, 20.5) \times 10^{-15} \, {\rm GeV} \, , \\[1ex]
&\varepsilon_K^{\rm SUSY} = -0.02 \, (-5.11, 4.92) \times 10^{-3} \, .
\end{align}
Note that the 95\% CI values for $\Delta M_K^{\rm SUSY}$ and $\varepsilon_K^{\rm SUSY}$ are larger than the uncertainties of those measurements, and some parameter sets potentially explain the discrepancy between the experimental result and SM prediction: for example, 870 samples can be $5.75 \times 10^{-4} \lesssim |\varepsilon_K^{\rm SUSY}| \lesssim 8.0 \times 10^{-4}$, which can compensate the gap between $|\varepsilon_K^{\rm exp}|$ and $|\varepsilon_K^{\rm SM}|_{\rm excl.}$, whose tension is $\sim 5\sigma$ level. 
This is one of the promising features of the model, as we do not need to take $M_{\rm SUSY}$ to be very heavy to suppress the mixing parameters. 
It is worth mentioning that the models with horizontal symmetries in Refs.~\cite{Leurer:1992wg,Nir:1993mx,Leurer:1993gy,Ibanez:1994ig,Nakai:2021mha}, the predicted values of $\Delta M_K$ and $\varepsilon_K$ tend to be smaller than those in our model, although these may be vary depending on the specific model setup. 
Combined with the prediction for $\Delta M_D$ discussed in section~\ref{sec:FCNC-upquark}, our model exhibits the pattern opposite to that of those models.  
Note that the SM predictions for these parameters calculated by \verb|susy_flavor_v2.54| are $\Delta M_K^{\rm SM} = 2.73 \times 10^{-15} \, {\rm GeV}$ and $\varepsilon_K^{\rm SM} = 2.34 \times 10^{-3}$. 
This $\varepsilon_K^{\rm SM}$ is larger than the SM prediction in Table~\ref{tablemeson}, because \verb|susy_flavor_v2.54| only calculates the short distance contributions. 
For more details about the calculation of $\varepsilon_K^{\rm SM}$, including not only short distance but also long distance contributions, see Refs.~\cite{Jwa:2025fon,Bailey:2015tba,Bailey:2018feb}.\footnote{According to Ref.~\cite{Jwa:2025fon}, sub-leading contributions from long distance effects will give about $-10\%$ deviation from the short distance result $\epsilon_K^{\rm SM}|_{\rm SD}$.}

Similar to the Kaon system, SUSY contributions to the $B_{d, s}^0$ meson mixing can be calculated as
\begin{align}
&| \Delta M^{\rm SUSY}_{B_d} | \simeq 1.3 \times 10^{-11} \times \left (\frac{3 \, {\rm TeV}}{M_{\rm SUSY}}\right)^2 \times \left| {\rm Re} \Bigl[ (\delta^d_{LL})_{13} (\delta^d_{RR})_{13} \Bigr] \right| \, {\rm GeV} \sim 8 \times 10^{-17} \, {\rm GeV} \, , \label{eq:DelMBdrough} \\
&| \Delta M^{\rm SUSY}_{B_s} | \simeq 1.9 \times 10^{-11} \times \left (\frac{3 \, {\rm TeV}}{M_{\rm SUSY}}\right)^2 \times \left| {\rm Re} \Bigl[ (\delta^d_{LL})_{23} (\delta^d_{RR})_{23} \Bigr] \right| \, {\rm GeV} \sim 6 \times 10^{-15} \, {\rm GeV} \, , \label{eq:DelMBsrough}
\end{align}
where we have used $\left| {\rm Re} \Bigl[ (\delta^d_{LL})_{13} (\delta^d_{RR})_{13} \Bigr] \right| \sim 6 \times 10^{-6}$ and $\left| {\rm Re} \Bigl[ (\delta^d_{LL})_{23} (\delta^d_{RR})_{23} \Bigr] \right| \sim 3 \times 10^{-4}$. 
These are similar to or smaller than the uncertainties of experimental results, and hence our model leads to smaller values for $\Delta M_{B_{d, s}}$ and is consistent with the experimental data, including the corresponding SM predictions in Table~\ref{tablemeson}. 
\begin{figure}[!tb]
\centering
\includegraphics[width=0.45\textwidth]{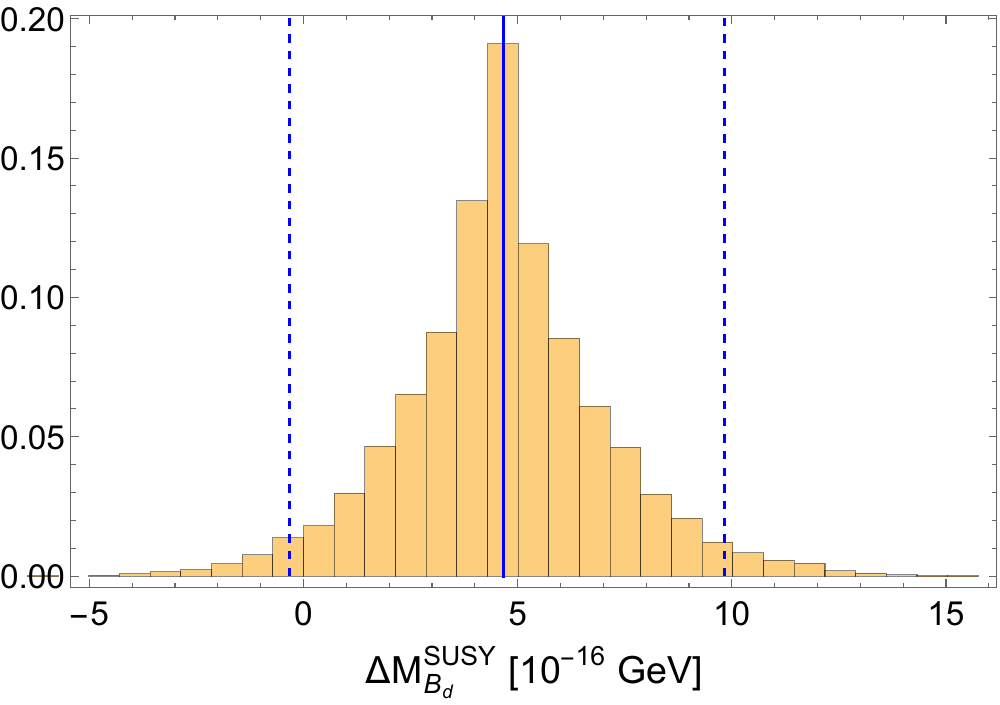} ~~ \includegraphics[width=0.45\textwidth]{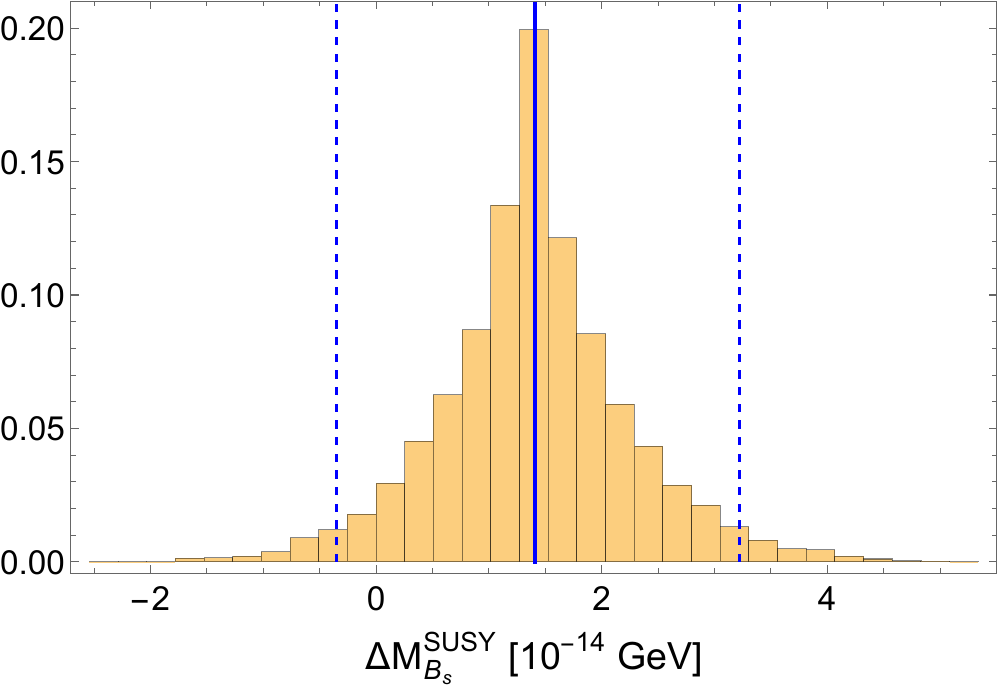}
\caption{Distributions for $\Delta M_{B_d}^{\rm SUSY}$ and $\Delta M_{B_s}^{\rm SUSY}$ in our model. 
The sold blue line corresponds to the mean value, and the 95\% CI is between two dashed blue lines. }
\label{fig:BBbar}
\end{figure}

In Fig.~\ref{fig:BBbar}, we show the distributions for $\Delta M_{B_d}^{\rm SUSY}$ (left panel) and $\Delta M_{B_s}^{\rm SUSY}$ (right panel), calculated via \verb|susy_flavor_v2.54|. 
The mean and 95\% CI values are:
\begin{align}
&\Delta M_{B_d}^{\rm SUSY} = 4.7 \, (-0.3, 9.8) \times 10^{-16} \, {\rm GeV} \, , \\[1ex]
&\Delta M_{B_s}^{\rm SUSY} = 1.4 \, (-0.4, 3.2) \times 10^{-14} \, {\rm GeV} \, .
\end{align}
These results for $\Delta M_{B_{d, s}}^{\rm SUSY}$ are almost consistent with our rough estimation in Eqs.~\eqref{eq:DelMBdrough} and \eqref{eq:DelMBsrough}, although mean values are slightly larger due to additional sub-leading contributions. 
Note that the SM predictions calculated by \verb|susy_flavor_v2.54| are $\Delta M_{B_d}^{\rm SM} = 3.51 \times 10^{-13} \, {\rm GeV}$ and $\Delta M_{B_s}^{\rm SM} = 1.20 \times 10^{-11}$ which are three orders of magnitude larger than our SUSY contributions, and these are consistent with those in Table~\ref{tablemeson}, because for $B$ meson system, the long distance contributions are sufficiently small. 

\subsubsection{Neutron EDM}

In addition to the CP-violating parameter in the Kaon system, our model leads to a nonzero neutron EDM. 
We expect that the new contribution should be small, because the phase in the CKM matrix is the only CP violation source. 
Therefore, even when $M_{\rm SUSY} = 3 \, {\rm TeV}$, the current upper limit on the neutron EDM, $1.8 \times 10^{-26} \, e \, {\rm cm}$ at 90\% CL~\cite{Abel:2020pzs}, is satisfied in our setup. 

The distribution for SUSY contributions to the neutron EDM is shown in Fig.~\ref{fig:dn}. 
\begin{figure}[!tb]
\centering
\includegraphics[width=0.45\textwidth]{Dist-dn.pdf}
\caption{Distribution for $|d_n|$ in our model. 
The sold blue line corresponds to the mean value, and the 95\% CI is between two dashed blue lines. }
\label{fig:dn}
\end{figure}
As expected, the size is one or two orders of magnitude smaller than the current experimental bound. 
The mean and 95\% CI values for $|d_n|$ are
\begin{align}
|d_n| = 2.35 \, (0.09, 6.47) \times 10^{-28} \, e \, {\rm cm} \, ,
\end{align}
and the typical value of $|d_n|$ in our setup with $M_{\rm SUSY} = 3 \, {\rm TeV}$ and $\tan \beta = 5$ is $|d_n| \sim 10^{-27} \mathchar`- 10^{-29} \, e \, {\rm cm}$. 
This result implies that for $M_{\rm SUSY} = 3 \, {\rm TeV}$ with $\tan \beta = 5$, new CP phases in the SUSY parameters are constrained to be similar or not much larger than the CP phase in the CKM matrix. 
Note that the n2EDM experiment aims to search for the neutron EDM with a sensitivity of $\sim 5 \times 10^{-28} \, e \, {\rm cm}$~\cite{Abel:2018yeo}, and therefore, some of our parameter space will be probed in the future. 

\subsubsection{$b \to s \gamma$}
\label{sec:bsgamma}

One of the most important and sensitive $\Delta F = 1$ processes for the down-type quark sector is $b \to s \gamma$. 
Note that in our model, the flavor-changing sources mainly come from the CKM matrix and the off-diagonal elements in $A_d$ which has similar hierarchy to $Y_d$. 
Hence, SUSY contributions are expected to be small, as we have observed for $\Delta M_{B_{d, s}}$ in section~\ref{sec:DelF=2}. 
In fact, FCNC processes for $B$ mesons are related to $(V_{\rm CKM})_{ub, td} \sim \lambda^3$ or $(V_{\rm CKM})_{cb, ts} \sim \lambda^2$, while those for the Kaon system are $(V_{\rm CKM})_{us, cd} \sim \lambda$, with $\lambda \approx 0.22$ being one of the Wolfenstein parameters in the CKM matrix~\cite{Wolfenstein:1983yz}. 
Our result for ${\rm BR} (b \to s \gamma)$ will demonstrate this feature. 
The dominant SUSY contribution to ${\rm BR} (b \to s \gamma)$, mediated by the gluino, is given by~\cite{Gabbiani:1996hi}
\begin{align}
{\rm BR} (b \to s \gamma)_{\rm SUSY} &\simeq \frac{10^{-7}}{{\rm GeV}^2} \times \left( \frac{3 \, {\rm TeV}}{M_{\rm SUSY}} \right)^4 \times \left( \left| \frac{1}{40} m_b (\delta^d_{LL})_{23} + \frac{1}{12} M_{\rm SUSY} (\delta^d_{LR})_{23} \right|^2 + L \leftrightarrow R \right) \nonumber \\[0.3ex]
&= 6.25 \times 10^{-3} \times \left( \frac{3 \, {\rm TeV}}{M_{\rm SUSY}} \right)^2 \Bigl[ |(\delta^d_{LR})_{23}|^2 + |(\delta^d_{LR})_{32}|^2 \Bigr] + \mathcal{O} \left( \frac{m_b^2}{M_{\rm SUSY}^2} \right) \, , \label{eq:b2sgam_SUSY}
\end{align}
where we have used the fact that $|(\delta^d_{RL})_{23}| = |(\delta^d_{LR})_{32}|$. 
Then, the rough estimation in our model is ${\rm BR} (b \to s \gamma)_{\rm SUSY} \lesssim 10^{-8}$, which is much smaller than the experimental result, ${\rm BR} (B \to X_s \gamma)_{\rm exp} = (3.49 \pm 0.19) \times 10^{-4}$~\cite{ParticleDataGroup:2024cfk,HeavyFlavorAveragingGroupHFLAV:2024ctg}. 
Note that the decay rate of $b \to s \gamma$ is the partonic one for $B \to X_s \gamma$ decay, and in principle, these decay rates can be different, due to non-perturbative effects. 
However, as studied in Refs.~\cite{Ali:1990tj,Falk:1993dh,Chetyrkin:1996vx,Benzke:2010js,Blake:2016olu,Paul:2016urs}, they are almost equivalent, $\Gamma (B \to X_s \gamma) \approx \Gamma (b \to s \gamma)$ with good accuracy.\footnote{Actually, $\Gamma (b \to s \gamma)$ matches with $\Gamma (B \to X_s \gamma)$ in the heavy $b$-quark mass limit, $m_b \to \infty$.}
Therefore, we use Eq.~\eqref{eq:b2sgam_SUSY} as our ${\rm BR} (B \to X_s \gamma)$ for checking the validity of our rough estimation with MI parameters. 

\begin{figure}[!tb]
\centering
\includegraphics[width=0.45\textwidth]{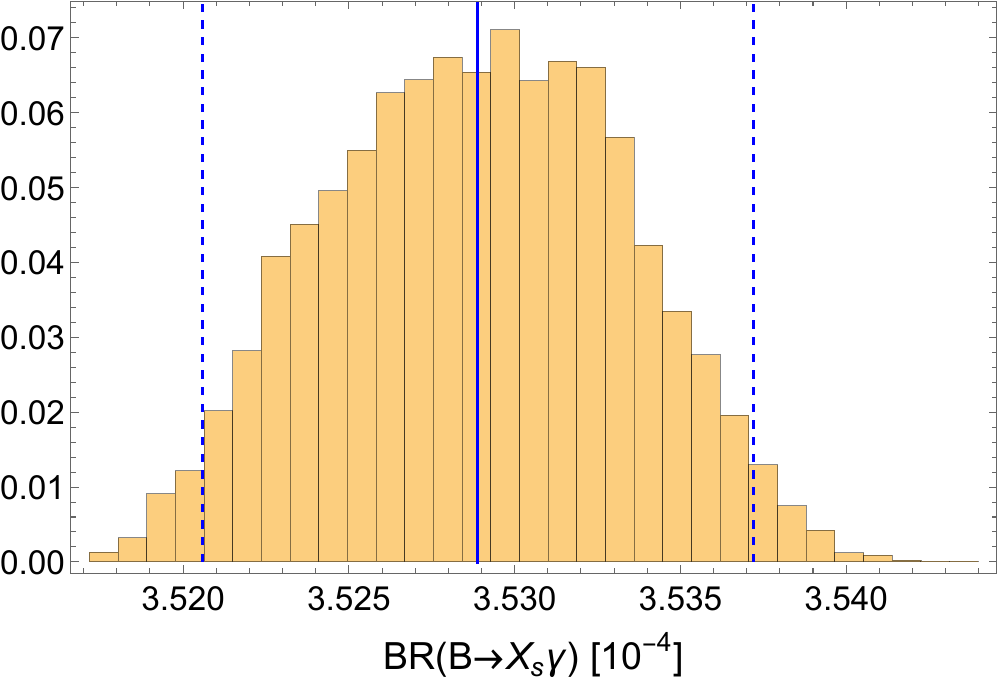}
\caption{Distribution for ${\rm BR} (B \to X_s \gamma)$ in our model. 
The sold blue line corresponds to the mean value, and the 95\% CI is between two dashed blue lines. }
\label{fig:B2Xsgam}
\end{figure}
The distribution for ${\rm BR} (B \to X_s \gamma)$ in our model is given in Fig.~\ref{fig:B2Xsgam}, which shows full results, namely ${\rm BR} (B \to X_s \gamma)_{\rm SM + SUSY}$. 
We then find
\begin{align}
{\rm BR} (B \to X_s \gamma)_{\rm SM + SUSY} = 3.529 \, (3.521, 3.537) \times 10^{-4} \, ,
\label{eq:B2Xsgam}
\end{align}
for the mean (95\% CI) value, and this is allowed by the current experimental result. 
The 95\% CI is narrow, which means that our new physics contributions are sufficiently small. 
In this case, the squared amplitude for $B \to X_s \gamma$ process can be estimated as
\begin{align}
|\mathcal{M}_{B \to X_s \gamma}|^2 = |\mathcal{M}_{B \to X_s \gamma}^{\rm SM} + \mathcal{M}_{B \to X_s \gamma}^{\rm SUSY}|^2 \simeq |\mathcal{M}_{B \to X_s \gamma}^{\rm SM}|^2 + 2 {\rm Re} \Big[ \mathcal{M}_{B \to X_s \gamma}^{\rm SM} \mathcal{M}_{B \to X_s \gamma}^{\rm SUSY \, *} \Bigr] \, ,
\label{eq:ampB2Xsgam}
\end{align}
neglecting the $|\mathcal{M}_{B \to X_s \gamma}^{\rm SUSY}|^2$ term. 
Then, it is expected that $|\mathcal{M}_{B \to X_s \gamma}^{\rm SM}|^2$ gives ${\rm BR} (B \to X_s \gamma) \sim 3.5 \times 10^{-4}$ as the SM prediction is calculated by \verb|susy_flavor_v2.54|, and we can consider that the distribution of ${\rm BR} (B \to X_s \gamma)$ in Fig.~\ref{fig:B2Xsgam} originates from the second term of the last expression in Eq.~\eqref{eq:ampB2Xsgam}. 
Then, according to 95\% CI in Eq.~\eqref{eq:B2Xsgam}, we can conclude that $|\mathcal{M}_{B \to X_s \gamma}^{\rm SUSY}| \sim 10^{-2} \times |\mathcal{M}_{B \to X_s \gamma}^{\rm SM}|$, and therefore, ${\rm BR} (B \to X_s \gamma)_{\rm SUSY} \sim 10^{-4} \times {\rm BR} (B \to X_s \gamma)_{\rm SM} \sim 10^{-8}$, which is consistent with our rough estimation by using MI parameters. 

\subsubsection{Other processes}
\label{sec:otherFCNCs}

Finally, we briefly summarize our SUSY contributions to other FCNC processes. 
For the Kaon system, $K^+ \to \pi^+ \nu \bar{\nu}$ and $K_L \to \pi^0 \nu \bar{\nu}$ may test and/or constrain the new physics contributions. 
In our setup, we obtain
\begin{align}
&{\rm BR} (K^+ \to \pi^+ \nu \bar{\nu}) = 9.18 \, (9.16, 9.21) \times 10^{-11} \, , \\[1ex]
&{\rm BR} (K_L \to \pi^0 \nu \bar{\nu}) = 3.2435 \, (3.2434, 3.2436) \times 10^{-11} \, ,
\end{align}
which are consistent with current experimental data~\cite{E949:2008btt,NA62:2021zjw,KOTO:2018dsc,KOTO:2020prk}. 
Here, the results are dominated by the SM contributions calculated via \verb|susy_flavor_v2.54|, and the sub-leading new physics contributions are given by the charged Higgs. 

The $B$ meson system has various processes of the leptonic and semi-leptonic decays. 
However, as we discussed for the $b \to s \gamma$ process in section~\ref{sec:bsgamma}, our SUSY contributions to these $B$ meson decays are expected to be small, and each deviation from the SM prediction will be too tiny to check signatures of our model at the current/future experiments. 
In fact, similar box diagrams for neutral $B$ meson oscillation with replacing two external quarks by charged leptons induce the processes of $B_{d, s} \to \ell \, \bar{\ell'}$. 
The contributions to LFV decays, $\ell \neq \ell'$, are extremely suppressed because of the lack of LFV sources in our setup, while we have the lepton flavor conserving processes, $\ell = \ell' (= e, \mu, \tau)$, whose mean values are obtained as follows:
\begin{align}
\begin{array}{|c|ccc|}
\hline
\ell & e & \mu & \tau \\ \hline
~ {\rm BR} (B_d \to \ell^+ \ell^-) ~ & ~ 2.91 \times 10^{-15} ~ & ~ 1.24 \times 10^{-10} ~ & ~ 2.60 \times 10^{-8} ~ \\
{\rm BR} (B_s \to \ell^+ \ell^-) & 9.84 \times 10^{-14} & 4.20 \times 10^{-9} & 8.92 \times 10^{-7} \\ \hline
\end{array}
\end{align}
The 95\% CIs are $- 0.077\% \sim 0.080\%$ (for $B_d$ decays) and $- 0.084\% \sim 0.086\%$ (for $B_s$ decays) of the corresponding mean values and independent of the lepton flavor. 
We have checked that all SUSY contributions are smaller than the mean values, as we expected. 
Experimentally, the processes with the muon in the final state are well-studied, and the current experimental results for these processes are~\cite{CMS:2022mgd,ATLAS:2018cur,CMS:2019bbr,LHCb:2021vsc}
\begin{align}
&{\rm BR} (B_d \to \mu^+ \mu^-) < 1.5 \times 10^{-10} ~~ (90\% ~ {\rm CL}) \, , \label{eq:Bd2mumuExp} \\[1ex]
&{\rm BR} (B_s \to \mu^+ \mu^-) = (3.34 \pm 0.27) \times 10^{-9} \, . \label{eq:Bs2mumuExp}
\end{align}
Comparing with these experimental results, our prediction for $B_d \to \mu^+ \mu^-$ is similar to the current upper bound, but smaller by a factor of $\sim 1.2$. 
On the other hand, for $B_s \to \mu^+ \mu^-$, we have an enhanced prediction which is outside of $3\sigma$ of the current experimental uncertainty in Eq.~\eqref{eq:Bs2mumuExp}. 
This enhancement of the mean value mainly originates from large CKM elements $(V_{\rm CKM})_{cb, ts}$. 
Note that experimental upper bounds for other decay processes are ${\rm BR} (B_d \to e^+ e^-) < 2.5 \times 10^{-9}$ and ${\rm BR} (B_s \to e^+ e^-) < 9.4 \times 10^{-9}$ for 90\% CL~\cite{LHCb:2020pcv} and ${\rm BR} (B_d \to \tau^+ \tau^-) < 2.1 \times 10^{-3}$ and ${\rm BR} (B_s \to \tau^+ \tau^-) < 6.8 \times 10^{-3}$ for 95\% CL~\cite{LHCb:2017myy}, and our predictions are 4-6 orders of magnitude smaller than these bounds. 

As the other $B$ meson decay processes, $B^+ \to \tau^+ \nu$ and $B \to D^{(*)} \tau \nu$ are also calculated by \verb|susy_flavor_v2.54|. 
However, the results are less interesting: ${\rm BR} (B^+ \to \tau^+ \nu) \simeq 8.8 \times 10^{-5}$ with 95\% CI of $\pm 10^{-3}\%$, and $R_{D^{(*)}} \equiv {\rm BR} (B \to D^{(*)} \tau \nu) / {\rm BR} (B \to D^{(*)} \ell \nu)$ for $\ell = e, \mu$ are almost fixed by the SM predictions. 
These results can be understood as follows. 
Since the main SUSY contributions for these processes come from the charged Higgs, and they are enhanced by large $\tan \beta$. 
The charged Higgs mass in our analysis is $m_{H^{\pm}} \simeq 4.6 \, (4.2, 5.1) \, {\rm TeV}$, and therefore, the contributions with $\tan \beta = 5$ cannot be large enough to overcome or be comparable to the SM predictions.

\section{Discussions}
\label{conclusion}

In the present paper, we have proposed a framework of the MSSM in which both the flavor structure and the suppression of FCNCs are controlled by non-invertible selection rules. 
By gauging the outer automorphism $\mathbb{Z}_2$ of a discrete $\mathbb{Z}_N$ symmetry, we obtained non-group-like fusion rules for matter fields that lead to realistic Yukawa textures reproducing the observed quark and lepton masses and mixings. 
A key outcome of this construction is that soft mass-squared matrices are forced to be diagonal, thereby suppressing dangerous FCNC processes. 
We evaluated the mass-insertion parameters with random $\mathcal{O}(1)$ coefficients and confirmed that flavor-violating effects in channels such as $\mu \to e \gamma$ and meson mixings remain safely below current experimental bounds. 
Remarkably, some parameter sets can explain the discrepancy of $\epsilon_K$ and the mass differences for $K$ and $B_{d,s}$ between their experimental results and SM predictions, even when the low SUSY-breaking scale $M_{\rm SUSY} = 3 \, {\rm TeV}$. 
The Yukawa textures and soft terms are sufficiently stable under renormalization group evolution, ensuring the robustness of the mechanism. 
Non-invertible selection rules constitute a powerful and novel approach to simultaneously explain realistic Yukawa textures and control FCNCs within the supersymmetric framework. 

It is natural to wonder whether our model can be embedded into the framework of grand unified theory (GUT).\footnote{For an attempt to construct $SU(5)$ GUT with non-invertible selection rules, see, Ref.~\cite{Kobayashi:2025rpx}.}
However, we can immediately observe that our Yukawa textures in Eq.~\eqref{eq:Yftexture} are not consistent with GUT, because common Yukawa structures among SM fermions are required due to matter unification. 
For instance, in a simple $SO(10)$ SUSY GUT model, we have only one Yukawa matrix for charged fermions.\footnote{For constructing a more robust model which is consistent with neutrino observables, we have additional Yukawa matrices with Higgs supermultiplets other than ${\bf 10}_H$, which are ${\bf 120}_H$ and $\overline{\bf 126}_H$. 
In this case, each Yukawa texture for a charged fermion at the GUT scale can get a more non-trivial form.}
This results in the same textures for all $Y_{u, d, e}$, and they should have off-diagonal elements to reproduce the CKM parameters. 
For soft mass-squared matrices, on the other hand, we can realize the same structures as those of our model: all of them have diagonal forms with $\mathcal{O} (1)$ ambiguity. 
The resultant MI parameters for the down-type quark sector will be at the same order of our $\delta_{LL, RR, LR}^d$, while sizable FCNCs in the up-type quark and charged lepton sectors will appear, and the model can be severely constrained by those observables, e.g., $D^0$--$\overline{D^0}$ mixing and $\mu \to e \gamma$. 
Detailed discussions in this direction will be performed in a future study. 

Alternatively, our model with non-invertible selection rules may be compatible with a higher dimensional orbifold GUT framework~\cite{Kawamura:2000ev,Hall:2001pg}, where a unified gauge group is broken by boundary conditions rather than large Higgs representations. 
We can expect that our non-invertible selection rules are incorporated into this structure by putting quark and lepton supermultiplets on a GUT breaking brane at an orbifold fixed point. 
Such a framework may provide a coherent ultraviolet picture linking flavor, SUSY and unification, and therefore deserves a further investigation. 

In addition to the FCNC problem which has been addressed in our framework, the MSSM generally suffers from the SUSY CP problem that generic soft SUSY breaking terms introduce new physical CP-violating phases and thus lead to unacceptably large EDMs, such as the electron EDM, unless the phases are tuned to be small (see e.g. Refs.~\cite{Pospelov:2005pr,Nakai:2016atk,Cesarotti:2018huy} and references therein). 
In alignment scenarios where flavor hierarchies are generated by horizontal symmetries, spontaneous CP violation (SCPV) driven by flavon fields can generate the observed CKM phase (for a discussion on the realization of SCPV in supersymmetric theories, see e.g. Ref.~\cite{Liu:2025ycm}), while new CP-violating phases are introduced to the soft terms in a controlled manner, suppressing SUSY contributions to CP-violating observables.\footnote{See also Ref.~\cite{Kobayashi:2025wty} for a CP-invariant theory with non-invertible selection rules.} 
In such scenarios, then, we can consider that non-invertible selection rules provide a complementary mechanism: they can restrict Yukawa structures and enforce diagonality (up to $\mathcal{O}(1)$ coefficients) in the sfermion mass-squared matrices and thereby reduce the number of physical CP-violating phases. 
A hybrid construction in which the Yukawa hierarchies originate from horizontal symmetries, while the pattern of allowed couplings is dictated by non-invertible selection rules, may achieve a more robust suppression of EDMs and FCNC processes while retaining realistic quark and lepton masses and mixings. 
We therefore expect that combining non-invertible selection rules with horizontal symmetries provides a promising avenue for a future work to address both the Yukawa hierarchies and the SUSY CP/FCNC problem in a unified manner.

\section*{Acknowledgments}

This work was supported by JSPS KAKENHI Grant Numbers JP25H01539 (H.O.) and JP26K07087 (H.O.). 
YN is supported by Natural Science Foundation of Shanghai. 
YS is supported by Natural Science Foundation of China under grant No. W2433006.

\appendix

\section{Various Yukawa textures for quarks}
\label{app}

Here, we list alternative Yukawa textures for quarks derived from the non-invertible selection rule $\tilde{\mathbb{Z}}_5^{(1)} \times \tilde{\mathbb{Z}}_5^{(2)}$. 
In particular, we focus on a three-zero texture for the down-quark mass matrix in the diagonal basis of the up-quark mass matrix.\footnote{Four or five zero textures for quarks proposed by Refs.~\cite{Fritzsch:1977vd,Fritzsch:1979zq,Ramond:1993kv} are not viable under the current experimental data.}
It has been pointed out in Ref.~\cite{Tanimoto:2016rqy} that there are six independent textures for the down-quark mass matrix that are consistent with experimental data. 
Note that Yukawa textures of the lepton sector can be taken as the same as those presented in the main text, thereby we arrive at the same conclusion for the lepton sector. 
Furthermore, the soft mass-squared matrices can be also diagonalized in the following assignments for matter superfields if the non-invertible selection rule is not violated in the SUSY breaking sector.

\subsection{$M_d^{(1)}$}

We consider the following assignments of matter superfields:
\begin{align}
Q_i &: ([g^2], [g^1], [g^2]) \, , \qquad U_i: ([g^1], [g^0], [g^1]) \, , \qquad D_i: ([g^0], [g^2], [g^2]) \, , \nonumber \\
H_u &: [g^1] \, , \qquad \qquad \qquad ~ H_d: [g^1] \, ,
\end{align}
for $\tilde{\mathbb{Z}}_5^{(1)}$ and
\begin{align}
Q_i &: ([g^2], [g^1], [g^1]) \, , \qquad U_i: ([g^1], [g^2], [g^0]) \, , \qquad D_i: ([g^2], [g^2], [g^0]) \, , \nonumber \\
H_u &: [g^1] \, , \qquad \qquad \qquad ~ H_d: [g^1] \, ,
\end{align}
for $\tilde{\mathbb{Z}}_5^{(2)}$. 
The selection rule of $\tilde{\mathbb{Z}}_5^{(1)} \times \tilde{\mathbb{Z}}_5^{(2)}$ leads to the following Yukawa textures:
\begin{align}
Y_{u} =
\begin{pmatrix}
* & 0 & 0 \\
0 & * & 0 \\
0 & 0 & *
\end{pmatrix} \, , \qquad Y_{d} =
\begin{pmatrix}
0 & * & 0 \\
* & * & * \\
0 & * & *
\end{pmatrix} \, ,
\end{align}
which correspond to $M_d^{(1)}$ in Ref.~\cite{Tanimoto:2016rqy}. 
This texture is also equivalent to $M_d^{(3)}$ under the unitary transformation of the right-handed quarks.

\subsection{$M_d^{(2)}$}

Let us consider the following assignments of matter superfields:
\begin{align}
Q_i &: ([g^2], [g^1], [g^2]) \, , \qquad U_i: ([g^1], [g^0], [g^1]) \, , \qquad D_i: ([g^2], [g^2], [g^2]) \, , \nonumber \\
H_u &: [g^1] \, , \qquad \qquad \qquad ~ H_d: [g^1] \, ,
\end{align}
for $\tilde{\mathbb{Z}}_5^{(1)}$ and
\begin{align}
Q_i &: ([g^2], [g^1], [g^1]) \, , \qquad U_i: ([g^1], [g^2], [g^0]) \, , \qquad D_i: ([g^1], [g^2], [g^0]) \, , \nonumber \\
H_u &: [g^1] \, , \qquad \qquad \qquad ~ H_d: [g^1] \, ,
\end{align}
for $\tilde{\mathbb{Z}}_5^{(2)}$. 
The selection rule of $\tilde{\mathbb{Z}}_5^{(1)} \times \tilde{\mathbb{Z}}_5^{(2)}$ gives
\begin{align}
Y_{u} =
\begin{pmatrix}
* & 0 & 0 \\
0 & * & 0 \\
0 & 0 & *
\end{pmatrix} \, , \qquad Y_{d} =
\begin{pmatrix}
* & * & 0 \\
0 & * & * \\
0 & * & *
\end{pmatrix} \, ,
\end{align}
which correspond to $M_d^{(2)}$ in Ref.~\cite{Tanimoto:2016rqy}. 
This texture is also equivalent to $M_d^{(16)}$ and $M_d^{(17)}$ under the unitary transformation of the right-handed quarks.

\subsection{$M_d^{(6)}$}

When we consider the following assignments of matter superfields:
\begin{align}
Q_i &: ([g^1], [g^1], [g^2]) \, , \qquad U_i: ([g^0], [g^0], [g^1]) \, , \qquad D_i: ([g^1], [g^0], [g^2]) \, , \nonumber \\
H_u &: [g^1] \, , \qquad \qquad \qquad ~ H_d: [g^1] \, ,
\end{align}
for $\tilde{\mathbb{Z}}_5^{(1)}$ and
\begin{align}
Q_i &: ([g^2], [g^1], [g^1]) \, , \qquad U_i: ([g^1], [g^0], [g^0]) \, , \qquad D_i: ([g^2], [g^2], [g^2]) \, , \nonumber \\
H_u &: [g^1] \, , \qquad \qquad \qquad ~ H_d: [g^1] \, ,
\end{align}
for $\tilde{\mathbb{Z}}_5^{(2)}$, the selection rule of $\tilde{\mathbb{Z}}_5^{(1)} \times \tilde{\mathbb{Z}}_5^{(2)}$ gives
\begin{align}
Y_{u} =
\begin{pmatrix}
* & 0 & 0 \\
0 & * & 0 \\
0 & 0 & *
\end{pmatrix} \, , \qquad Y_{d} =
\begin{pmatrix}
0 & * & * \\
0 & * & * \\
* & 0 & *
\end{pmatrix} \, ,
\end{align}
which correspond to $M_d^{(6)}$ in Ref.~\cite{Tanimoto:2016rqy}.

\subsection{$M_d^{(12)}$}

When we consider the following assignments of matter superfields:
\begin{align}
Q_i &: ([g^1], [g^2], [g^1]) \, , \qquad U_i: ([g^0], [g^1], [g^0]) \, , \qquad D_i: ([g^1], [g^0], [g^2]) \, , \nonumber \\
H_u &: [g^1] \, , \qquad \qquad \qquad ~ H_d: [g^1] \, ,
\end{align}
for $\tilde{\mathbb{Z}}_5^{(1)}$ and
\begin{align}
Q_i &: ([g^2], [g^2], [g^1]) \, , \qquad U_i: ([g^1], [g^1], [g^0]) \, , \qquad D_i: ([g^1], [g^2], [g^2]) \, , \nonumber \\
H_u &: [g^1] \, , \qquad \qquad \qquad ~ H_d: [g^1] \, ,
\end{align}
for $\tilde{\mathbb{Z}}_5^{(2)}$, the selection rule of $\tilde{\mathbb{Z}}_5^{(1)} \times \tilde{\mathbb{Z}}_5^{(2)}$ gives
\begin{align}
Y_{u} =
\begin{pmatrix}
* & 0 & 0 \\
0 & * & 0 \\
0 & 0 & *
\end{pmatrix} \, , \qquad Y_{d} =
\begin{pmatrix}
0 & * & * \\
* & 0 & * \\
0 & * & *
\end{pmatrix} \, ,
\end{align}
which correspond to $M_d^{(12)}$ in Ref.~\cite{Tanimoto:2016rqy}.

\subsection{$M_d^{(13)}$}

When we consider the following assignments of matter superfields:
\begin{align}
Q_i &: ([g^2], [g^1], [g^2]) \, , \qquad U_i: ([g^1], [g^0], [g^1]) \, , \qquad D_i: ([g^1], [g^1], [g^2]) \, , \nonumber \\
H_u &: [g^1] \, , \qquad \qquad \qquad ~ H_d: [g^1] \, ,
\end{align}
for $\tilde{\mathbb{Z}}_5^{(1)}$ and
\begin{align}
Q_i &: ([g^2], [g^1], [g^1]) \, , \qquad U_i: ([g^1], [g^2], [g^0]) \, , \qquad D_i: ([g^0], [g^2], [g^2]) \, , \nonumber \\
H_u &: [g^1] \, , \qquad \qquad \qquad ~ H_d: [g^1] \, ,
\end{align}
for $\tilde{\mathbb{Z}}_5^{(2)}$, the selection rule of $\tilde{\mathbb{Z}}_5^{(1)} \times \tilde{\mathbb{Z}}_5^{(2)}$ gives
\begin{align}
Y_{u} =
\begin{pmatrix}
* & 0 & 0 \\
0 & * & 0 \\
0 & 0 & *
\end{pmatrix} \, , \qquad Y_{d} =
\begin{pmatrix}
0 & * & * \\
0 & 0 & * \\
* & * & *
\end{pmatrix} \, ,
\end{align}
which correspond to $M_d^{(13)}$ in Ref.~\cite{Tanimoto:2016rqy}. 
This texture is also equivalent to $M_d^{(11)}$ and $M_d^{(15)}$ under the unitary transformation of the right-handed quarks.

\subsection{$M_d^{(14)}$}

When we consider the following assignments of matter superfields:
\begin{align}
Q_i &: ([g^2], [g^1], [g^1]) \, , \qquad U_i: ([g^1], [g^0], [g^0]) \, , \qquad D_i: ([g^2], [g^1], [g^2]) \, , \nonumber \\
H_u &: [g^1] \, , \qquad \qquad \qquad ~ H_d: [g^1] \, ,
\end{align}
for $\tilde{\mathbb{Z}}_5^{(1)}$ and
\begin{align}
Q_i &: ([g^2], [g^2], [g^1]) \, , \qquad U_i: ([g^1], [g^1], [g^0]) \, , \qquad D_i: ([g^1], [g^2], [g^2]) \, , \nonumber \\
H_u &: [g^1] \, , \qquad \qquad \qquad ~ H_d: [g^1] \, ,
\end{align}
for $\tilde{\mathbb{Z}}_5^{(2)}$, the selection rule of $\tilde{\mathbb{Z}}_5^{(1)} \times \tilde{\mathbb{Z}}_5^{(2)}$ gives
\begin{align}
Y_{u} =
\begin{pmatrix}
* & 0 & 0 \\
0 & * & 0 \\
0 & 0 & *
\end{pmatrix} \, , \qquad Y_{d} =
\begin{pmatrix}
* & * & * \\
* & 0 & * \\
0 & 0 & *
\end{pmatrix} \, ,
\end{align}
which correspond to $M_d^{(14)}$ in Ref.~\cite{Tanimoto:2016rqy}. 
This texture is also equivalent to $M_d^{(4)}$ and $M_d^{(5)}$ under the unitary transformation of the right-handed quarks.

\bibliography{ref}{}
\bibliographystyle{JHEP} 

\end{document}